\documentclass[preprint,superscriptaddress,showkeys]{revtex4}

\usepackage{graphics}
\usepackage[dvipdfmx]{graphicx}
\usepackage{amssymb}
\usepackage{amsmath}
\usepackage{amsfonts}
\usepackage{setspace}

\newif\ifFIG
\FIGtrue

\begin{document}

\title{Theoretical study of lithium ionic conductors by electronic stress tensor density and  electronic kinetic energy density}

\author{Hiroo Nozaki}
\author{Yosuke Fujii}
\author{Kazuhide Ichikawa}
%\affiliation{Department of Micro Engineering, Kyoto University, Kyoto 615-8540, Japan}
\affiliation{Department of Micro Engineering, Kyoto University, Bldg.~C3, Kyotodaigakukatsura, Nishikyo-ku, Kyoto-shi,
Kyoto, 615-8540, Japan}

\author{Taku Watanabe} 
\author{Yuichi Aihara} 
%\affiliation{Samsung R\&D Institute Japan, Minoh-shi, Osaka 562-0036, Japan}
%\affiliation{Samsung R\&D Institute Japan, Minoh Semba Center Bldg., 13F, Semba Nishi 2-1-11, Minoh-shi, Osaka, 562-0036, Japan}
\affiliation{AR Center, Samsung R\&D Institute Japan, Minoh Semba Center Bldg., Semba Nishi 2-1-11, Minoh, Osaka 562-0036, Japan}

\author{Akitomo Tachibana} 
\email{akitomo@scl.kyoto-u.ac.jp}
%\affiliation{Department of Micro Engineering, Kyoto University, Kyoto 615-8540, Japan}
\affiliation{Department of Micro Engineering, Kyoto University, Bldg.~C3, Kyotodaigakukatsura, Nishikyo-ku, Kyoto-shi,
Kyoto, 615-8540, Japan}

\date{\today}

\begin{abstract}
We analyze the electronic structure of lithium ionic conductors, ${\rm Li_3PO_4}$ and ${\rm Li_3PS_4}$, using the electronic stress tensor density and kinetic energy density with special focus on the ionic bonds among them.
We find that, as long as we examine the pattern of the eigenvalues of the electronic stress tensor density, we cannot distinguish between the ionic bonds and bonds among metalloid atoms. 
We then show that they can be distinguished by looking at the morphology of the electronic interface, the zero surface of the electronic kinetic energy density. 
\end{abstract}

  %revtex

\maketitle

%%%%%%%%%%%%%%%%%%%%%%%%%%%%
\section{Introduction} \label{sec:intro}
%%%%%%%%%%%%%%%%%%%%%%%%%%%%

Lithium-ion batteries are indispensable for our daily life and further improvements are pursued. 
One of the promising ways is the conversion to the solid-state batteries, and effective lithium ionic conductors for the solid electrolytes are now being searched for \cite{Kamaya2011,Wang2015}. 
The solid electrolytes have several advantages over the liquid ones such as better chemical and physical stability. However, there are also disadvantages, for example lower ionic conductivity, which should be overcome by new materials.
In order for efficient material search, it is important to characterize the chemical bonding of the lithium ionic conductors through the quantum chemical electronic structure calculation, and establish the connection with their material properties like chemical stability, diffusivity and so on.
We have been developing the characterization scheme of electronic structure using the electronic stress tensor density and kinetic energy density based on the rigged quantum electrodynamics (RQED) theory \cite{Tachibana2001, Tachibana2003, Tachibana2010, Tachibana2013, Tachibana2014a},
and have applied it to various quantum systems \cite{Tachibana2001,Ikenaga2002, Hotta2002, Hasegawa2003, Yoshida2003, Makita2003, Tachibana2003, Tachibana2004,Kawakami2004, Tachibana2005, Doi2005, Nakamura2005, Nakano2006, Szarek2007, Szarek2008, Fukushima2008, Szarek2009, Ichikawa2009a, Ichikawa2009b,Tachibana2010, Ichikawa2010, Fukushima2010, Ichikawa2011a, Ichikawa2011b, Ichikawa2011c, Fukushima2011, Ikeda2011, Senami2011, Henry2011,Ichikawa2012,Tachibana2013,Tachibana2014a, Nozaki2015,Nozaki2015b}. 
As a preliminary stage of our research, we have applied our method to the crystal structures of ${\rm Li_3PO_4}$ and ${\rm Li_3PS_4}$.
In particular, the latter serves as a prototype for sulfide solid electrolytes which have been intensively studied recently \cite{Hayamizu2013,Agostini2013,Hayamizu2014,Ito2014,Yamada2015,Hayamizu2015}.
In the course of this study, we have found a method to characterize ionic bonding by the electronic stress tensor density and kinetic energy density. 
In this paper, we wish to discuss how the ionicity of chemical bonding is expressed using these quantities. 

In our previous works, we have studied how we may characterize covalency and metallicity of chemical bonding in view of the electronic stress tensor density. 
First, it has been proposed in Ref.~\cite{Tachibana2004} that the ``spindle structure", where the largest eigenvalue of the electronic stress tensor is positive and the corresponding eigenvectors form a bundle of flow lines that connects nuclei, can characterize and visualize the bonding region with covalency.
Then, we have proposed that the negativity of the three eigenvalues of the stress tensor and their degeneracy,
which is the same pattern as liquid, may characterize some aspects of the metallicity of chemical bonding \cite{Szarek2007,Ichikawa2012,Tachibana2013}. 
Specifically, in Ref.~\cite{Ichikawa2012}, we have found that the three eigenvalues of the Li and Na clusters have almost same values while 
the hydrocarbon molecules have the largest eigenvalue much larger than the second largest eigenvalue, 
which has similar value to the smallest eigenvalue. 
The former degeneracy pattern suggests that the bonds are not directional while the latter implies the clear directionality of the bonds,
respectively reflecting the metallicity and covalency of chemical bonding. 
Furthermore, recently, we have found in Ref.~\cite{Nozaki2015} that  the chemical bonds between Ge, Sb and Te atoms, which are usually classified as metalloids, exhibit intermediate properties between alkali metals and hydrocarbon molecules in terms of the sign and degeneracy pattern of the eigenvalues.

Based on these findings, it is now worth asking how the ionicity of chemical bonding is discriminated from the covalency and metallicity from the viewpoint of the electronic stress tensor density. 
Actually, as will be shown below, the stress tensor alone is not sufficient to characterize the ionicity and we need to see the pattern of the electronic kinetic energy density.
Our definition of the kinetic energy density is not positive-definite and there exist zero surfaces, which are designated as ``electronic interfaces" \cite{Tachibana2001}.
 The outermost electronic interface can give a clear image of the intrinsic shape of atoms and molecules, and it has been used to investigate various chemical reactions in our past works \cite{Ikenaga2002,Hotta2002,Hasegawa2003,Yoshida2003,Makita2003,Kawakami2004,Tachibana2004,Doi2005,Nakamura2005,Nakano2006,Fukushima2008,Szarek2009,Fukushima2010,Fukushima2011,Ikeda2011,Senami2011,Henry2011,Ichikawa2012}.
To characterize ionicity, the electronic interface is found to play an important role.

This paper is organized as follows.
In Sec.~\ref{sec:theory}, we briefly review our method to analyze electronic structures using the electronic stress tensor density and kinetic energy density, including their definitions. 
In Sec.~\ref{sec:results}, we first describe our data set and computational details. We then show the eigenvalues and eigenvectors of the electronic stress tensor of the lithium ionic conductors. We also show their electronic interfaces and how they can be used to characterize the ionicity of the chemical bonding. 
The final section is devoted to our conclusion.

%%%%%%%%%%%%%%%%%%%%%%%%%%%%
\section{Theory} \label{sec:theory}
%%%%%%%%%%%%%%%%%%%%%%%%%%%%

In this paper, we analyze the electronic structure of molecules using quantities based on the RQED theory \cite{Tachibana2001, Tachibana2003, Tachibana2010, Tachibana2013, Tachibana2014a}.
Due to the field theoretic nature, the theory includes local quantities defined at each point in space which are useful to describe quantum systems, such as the electronic stress tensor density and the kinetic energy density.
We briefly review them in this section. 

We here summarize our notations. $\hat{\psi}(x)$ is the four-component Dirac field operator for electrons, and $\hat{A}_k(x)$ is the vector potential of the photon field operator in the Coulomb gauge (${\rm div} \hat{\vec{A}}(x)=0$). 
The spacetime coordinate is expressed as $x=(ct, \vec{r})$ where $c$ denotes the speed of light in vacuum.
$\hbar$ is the reduced Planck constant, $e$ is the electron charge magnitude ($e > 0$), $m$ is the electron mass, and $\gamma^\mu$ ($\mu=$0-3) are the gamma matrices. 
The dagger as a superscript is used to express Hermite conjugate, and $\hat{\bar{\psi}}(x) \equiv \hat{\psi}^\dagger(x) \gamma^0$.
The Latin letter indices like $k$ and $l$ express space coordinates from 1 to 3, and repeated indices 
implies a summation over 1 to 3. 

%%%%%%%%%%%%%%%%%%%%%%%%%%%%%%%%%%%%%%%%%%%
\subsection{Electronic stress tensor density and tension density} \label{sec:stress}
%%%%%%%%%%%%%%%%%%%%%%%%%%%%%%%%%%%%%%%%%%%
The electronic stress tensor density operator $\hat{\tau}^{\Pi\, kl}_e(x)$ is defined as 
\begin{eqnarray}
\hat{\tau}^{\Pi\, kl}_e(x) = \frac{i\hbar c}{2} 
\left[
\hat{\bar{\psi}}(x) \gamma^l \hat{D}_{e\, k}(x) \hat{\psi}(x) 
- \left( \hat{D}_{e\, k}(x) \hat{\psi}(x)\right)^\dagger \gamma^0 \gamma^l \hat{\psi}(x)
\right],   \label{eq:op_stress}
\end{eqnarray}
where $\hat{D}_{e\, k}(x) = \partial_k + i\frac{Z_e e}{\hbar c} \hat{A}_k(x)$ with $Z_e = -1$ is the gauge covariant derivative \cite{Tachibana2001}.
We note that the stress tensor is not defined uniquely since mathematically any tensor whose divergence is zero can be added to.
We adopt  Eq.~\eqref{eq:op_stress}, since this is a minimal combination respecting the Lorentz covariance, gauge invariance and hermiticity. 

The equation of motion for the electronic kinetic momentum density operator 
$\hat{\vec{\Pi}}_e(x) = \frac{1}{2} \left(
i\hbar \hat{\psi}^\dagger(x) \hat{\vec{D}}_e(x) \hat{\psi}(x) - i\hbar \left( \hat{\vec{D}}_e(x)\hat{\psi}(x) \right)^\dagger \cdot \hat{\psi}(x) \right)$
is expressed using $\hat{\tau}^{\Pi\, kl}_e(x)$.
Namely, the time derivative of $\hat{\vec{\Pi}}_e(x)$ equals to the sum of the Lorentz force density operator $\hat{\vec{L}}_e(x)$ and the tension density operator $\hat{\vec{\tau}}^\Pi_e(x)$,
which is the divergence of $\hat{\tau}^{\Pi\, kl}_e(x)$:
\begin{eqnarray}
\frac{\partial}{\partial t} \hat{\vec{\Pi}}_e(x) &=& \hat{\vec{L}}_e(x) + \hat{\vec{\tau}}^\Pi_e(x),   \label{eq:dPiedt} \\
\hat{\vec{L}}_e(x) &=& \hat{\vec{E}}(x) \hat{\rho}_e(x) + \frac{1}{c} \hat{\vec{j}}_e(x) \times \hat{\vec{B}}(x),  \label{eq:op_Lorentz}  \\
\hat{\tau}^{\Pi k}_e(x) &=& \partial_l \hat{\tau}^{\Pi\, kl}_e(x) \\
&=&
\frac{i\hbar c}{2} \Bigg[ \left(\hat{D}_{el}(x)\hat{\psi}(x) \right)^\dagger \gamma^0 \gamma^l \cdot \hat{D}_{ek}(x) \hat{\psi}(x)
+ \hat{\bar{\psi}}(x) \gamma^l \hat{D}_{ek}(x) \hat{D}_{el}(x) \hat{\psi}(x)  \nonumber \\
& &-\left( \hat{D}_{ek}(x)\hat{D}_{el}(x)\hat{\psi}(x) \right)^\dagger \gamma^0 \gamma^l \cdot \hat{\psi}(x)
-\left( \hat{D}_{ek}(x)\hat{\psi}(x) \right)^\dagger \gamma^0 \gamma^l \cdot \hat{D}_{el}(x) \hat{\psi}(x) \Bigg]  \nonumber \\
& &
-\frac{1}{c} \left( \hat{\vec{j}}_e(x) \times \vec{B}(x) \right)^k .  \label{eq:op_tension}
\end{eqnarray}
In these equations, $\hat{\rho}_e(x)$, $\hat{\vec{j}}_e(x)$, $\hat{\vec{E}}(x)$ and $\hat{\vec{B}}(x)$ are the electronic charge density operator, electronic charge current density operator, electric field operator, and magnetic field operator respectively.

In this paper, as we study nonrelativistic systems, we approximate the expressions above in the framework of the primary RQED \cite{Tachibana2013,Tachibana2014a}, in which the small components of the four-component electron field are expressed by the large components as $\hat{\psi}_S (x) \approx  - \frac{1}{2m c} i\hbar \sigma^k  \hat{D}_{ek}(x) \hat{\psi}_L (x)$ and the spin-dependent terms are ignored. 
With this approximation, Eq.~\eqref{eq:op_stress} can be expressed as
\begin{eqnarray}
\hat{\tau}^{\Pi\, kl}_e(x) &\approx& 
\frac{\hbar^2}{4m} \bigg[ \hat{\psi}^\dagger_L(x) \hat{D}_{ek}(x) \hat{D}_{el}(x)  \hat{\psi}_L(x)
+ \left(\hat{D}_{ek}(x)\hat{D}_{el}(x)  \hat{\psi}_L(x) \right)^\dagger  \cdot \hat{\psi}_L(x) \nonumber \\
& & - \left( \hat{D}_{ek}(x)  \hat{\psi}_L(x) \right)^\dagger \cdot  \hat{D}_{el}(x)  \hat{\psi}_L(x)
- \left( \hat{D}_{el}(x)  \hat{\psi}_L(x) \right)^\dagger \cdot  \hat{D}_{ek}(x)  \hat{\psi}_L(x) \bigg] \nonumber \\
& \equiv& \hat{\tau}^{S\, kl}_e(x),  \label{eq:op_stressPR}
\end{eqnarray}
and, $\hat{\vec{\Pi}}_e(x) \approx m \hat{\vec{v}}_e(x)$, where the velocity density operator is
\begin{eqnarray}
\hat{\vec{v}}_e(x) = \frac{i\hbar}{2m} \left[ \hat{\psi}^\dagger_L(x) \hat{\vec{D}}_{e}(x) \hat{\psi}_L(x) 
-   \left( \hat{\vec{D}}_{e}(x) \hat{\psi}_L(x) \right)^\dagger \cdot  \hat{\psi}_L(x)  \right].
\end{eqnarray}
Its equation of motion is
\begin{eqnarray}
\frac{\partial}{\partial t} \left( m \hat{\vec{v}}_e(x) \right) = \hat{\vec{L}}_e(x) + \hat{\vec{\tau}}^S_e(x),   \label{eq:dmvedt}
\end{eqnarray}
where
\begin{eqnarray}
\hat{\tau}^{S k}_e(x) &=& \partial_l \hat{\tau}^{S\, kl}_e(x) \\
&=&
\frac{\hbar^2}{4m} \bigg[  \hat{\psi}^\dagger_L(x) \hat{D}_{ek}(x) \hat{\vec{D}}^2_{e}(x)  \hat{\psi}_L(x)
+ \left(\hat{D}_{ek}(x)\hat{\vec{D}}^2_{e}(x)  \hat{\psi}_L(x) \right)^\dagger  \cdot \hat{\psi}_L(x) \nonumber \\
& & - \left( \hat{D}_{ek}(x)  \hat{\psi}_L(x) \right)^\dagger \cdot  \hat{\vec{D}}^2_{e}(x)  \hat{\psi}_L(x)
- \left( \hat{\vec{D}}^2_{e}(x)  \hat{\psi}_L(x) \right)^\dagger \cdot  \hat{D}_{ek}(x)  \hat{\psi}_L(x) \bigg] \nonumber \\
& & -\frac{1}{c} \left( \hat{\vec{j}}_e(x) \times \vec{B}(x) \right)^k,   \label{eq:op_tensionPR}
\end{eqnarray}
which is also obtained from the primary RQED approximation of Eq.~\eqref{eq:op_tension}.
As for the Lorentz force density operator, $\hat{\rho}_e(x) = \hat{\psi}^\dagger_L(x) \hat{\psi}_L(x)$ and $\hat{\vec{j}}_e(x) =  Z_e e \hat{\vec{v}}_e(x)$ are used
in Eq.~\eqref{eq:op_Lorentz}. 
 $\hat{\psi}_L(x)$ is regarded as the two-component spinor operator representing non-relativistic electronic field (which is denoted by $\hat{\chi}$ in Ref.~\cite{Tachibana2001}, and Eqs.~\eqref{eq:op_stressPR} and \eqref{eq:op_tensionPR} respectively coincide with Eqs.~(12) and (11) in that paper).
Then, taking the expectation value of %Eq.~\eqref{eq:dPiedt} with respect to the stationary state of the electrostatic Hamiltonian, we obtain 
Eq.~\eqref{eq:dmvedt} with respect to the stationary state of the electrostatic Hamiltonian, we obtain 
\begin{eqnarray}
0 = \langle \hat{L}^k_e(x) \rangle + \langle \hat{\tau}^{S k}_e(x) \rangle= \langle \hat{L}^k_e(x) \rangle + \partial_l \langle  \hat{\tau}^{S\, kl}_e(x) \rangle,	\label{eq:equilibrium}
\end{eqnarray}
which shows the balance between electromagnetic force and tension at each point in space. 
As Eq.~\eqref{eq:equilibrium} has a form of the equilibrium equation that the tension keeps the electrons in the stationary bound state in atomic and molecular systems, the stress tensor density and tension density carry fundamental information of these systems.
%We note that they are derived based on the quantum field theoretic consideration.
We note that, although they are derived based on the quantum field theoretic consideration, their expectation values are expressed in terms of the wave function of the system we study, which is obtained by solving the non-relativistic time-independent Schr\"{o}dinger equation.
For simplicity, we denote $\langle \hat{\tau}_e^{S k}(x) \rangle$ and $\langle \hat{\tau}_e^{Skl}(x) \rangle$ respectively as $\tau_e^{Sk}(\vec{r})$ and $\tau_e^{Skl}(\vec{r})$, which are explicitly written as
\begin{eqnarray} 
\tau^{Skl}_{e}(\vec{r}) &=& \frac{\hbar^2}{4m}\sum_i \nu_i
\Bigg[\psi^*_i(\vec{r})\frac{\partial^2\psi_i(\vec{r})}{\partial x^k \partial x^l}-\frac{\partial\psi^*_i(\vec{r})}{\partial x^k} \frac{\partial\psi_i(\vec{r})}{\partial x^l} \nonumber\\
& & \hspace{4cm} +\frac{\partial^2 \psi^*_i(\vec{r})}{\partial x^k \partial x^l}\psi_i(\vec{r}) -\frac{\partial \psi^*_i(\vec{r})}{\partial x^l}\frac{\partial \psi_i(\vec{r})}{\partial x^k}\Bigg], \label{eq:stress}
\end{eqnarray}
\begin{eqnarray} 
\tau^{S k}_{e}(\vec{r}) &=&  \partial_l  \tau_e^{Skl}(\vec{r}) \nonumber \\
&=&\frac{\hbar^2}{4m}\sum_i \nu_i
\Bigg[\psi^*_i(\vec{r})\frac{\partial \Delta\psi_i(\vec{r})}{\partial x^k}-\frac{\partial\psi^*_i(\vec{r})}{\partial x^k} \Delta\psi_i(\vec{r}) \nonumber\\
& & \hspace{4cm} +\frac{\partial \Delta\psi^*_i(\vec{r})}{\partial x^k}\psi_i(\vec{r}) -\Delta \psi^*_i(\vec{r}) \frac{\partial \psi_i(\vec{r})}{\partial x^k}\Bigg],
\label{eq:tension}
\end{eqnarray}
where $\psi_i(\vec{r})$ is the $i$th natural orbital, $\nu_i$ is its occupation number, and
$\Delta$ denotes the Laplacian, $\Delta \equiv \sum_{k=1}^3 (\partial/\partial x^k)^2$.
In these expressions, as we consider stationary state, we write only spatial coordinate $\vec{r}$. 
The eigenvalue of the symmetric tensor $\stackrel{\leftrightarrow}{\tau}_e^{S}$ is the principal stress and the eigenvector is the principal axis. We define the ordering of eigenvalues as $\tau^{S11}_{e}(\vec{r}) \le \tau^{S22}_{e}(\vec{r}) \le \tau^{S33}_{e}(\vec{r})$.
It may be worth explaining here the stress tensor in somewhat intuitive manner. 
If a straight solid bar is pulled from its both ends, the stress is called tensile
with respect to the bar's cross section (which is perpendicular to the direction of the bar).
Namely, a part of the bar on one side of the cross section pulls up a part on the other side. 
For the opposite case, if it is pushed, the stress is called compressive.
In this case, a part of the bar on one side of the cross section pushes back a part on the other side. 
In this analogy, the direction of the bar corresponds to the direction of the eigenvector,
and the eigenvalue tells us whether the stress is tensile (positive eigenvalue) or 
compressive (negative eigenvalue). 
Similarly, what the electronic stress tensor density tells us is the pattern of interaction
of the electronic field at each point in space with the neighboring electronic field. 
You may imagine a fictitious plane normal to the eigenvector at a point in atomic or molecular system and if the corresponding eigenvalue is 
positive (negative), the electronic field on one side of the plane pulls up (pushes back) the electron field on the other side.

We note in passing that relativistic expression for Eqs.~\eqref{eq:stress} and \eqref{eq:tension} may be derived in a straightforward manner from Eq.~\eqref{eq:op_stress} and \eqref{eq:op_tension} using four-component relativistic wavefunctions. It is, however, problematic to derive the expressions using two-component relativistic wavefunctions as is pointed out in Ref.~\cite{Fukuda2016}. It has been argued  that the local physical quantities like stress tensor density and tension density
cannot be derived rigorously by the two-component wave function defined by
the Foldy-Wouthuysen-Tani transformation with the existence of the
Dirac mass term.

We define the Lagrange point ${\vec r}_L$ as the point between an atomic pair at which $\vec{\tau}^S(\vec{r})$ is a zero vector ($\tau^{S k}(\vec{r}_L)=0$). This is originally proposed in Ref.~\cite{Szarek2007}. We use it as a point which can characterize a bond between two atoms. 
In this paper, we search a local minimum in the region between two atoms and if $|\vec{\tau}^S(\vec{r})|$ at that point is below $1.0 \times 10^{-3}$\,a.u., we regard the point as the Lagrange point. 
We report the eigenvalues of electronic stress tensor density at this point in the following sections.

We refer Refs.~\cite{Epstein1975, Bader1980, Bamzai1981a, Nielsen1983, Nielsen1985, Folland1986a, Folland1986b, Godfrey1988, Filippetti2000, Pendas2002, Rogers2002, Morante2006, Tao2008, Ayers2009, Jenkins2011,GuevaraGarcia2011,Finzel2013,Finzel2014a}
for other studies of quantum systems with the stress tensor in somewhat different contexts and definitions.
Recently, the difference in the definitions and approximations has been discussed in Refs.~\cite{Ayers2009, Jenkins2011, GuevaraGarcia2011,Finzel2013,Finzel2014a}.
Our stress tensor density Eq.~\eqref{eq:stress} is same as the one in Ref.~\cite{Bader1980}, Eq.~(22), although our definition originates from the quantum field theoretic consideration based on RQED. 
We advocate the use of Eq.~\eqref{eq:stress} since it comes from the stress tensor density operator Eq.~\eqref{eq:op_stress}, which minimally respects reasonable physical principles.  
Moreover, this definition turns out to be phenomenologically useful as shown by our works mentioned in the previous section.
We also note that the tension density in the form of Eq.~\eqref{eq:tension} is same as what is called quantum force density in Refs.~\cite{Pendas2002,Pendas2012}. Then, in the stationary state, the Ehrenfest force field used in Refs.~\cite{Pendas2002,Pendas2012,Maza2013,Dillen2015} (and the force density in Ref.~\cite{Bader1980}, Eq.~(24)) only differs from the tension density by  the minus sign.

%%%%%%%%%%%%%%%%%%%%%%%%%%%%%%%%%%%%%%%%%%%
\subsection{Electronic kinetic energy density} \label{sec:ked}
%%%%%%%%%%%%%%%%%%%%%%%%%%%%%%%%%%%%%%%%%%%
The electronic kinetic energy density operator $\hat{T}_e(x)$ is defined as \cite{Tachibana2001}
\begin{eqnarray}
\hat{T}_e(x) &=& -\frac{\hbar^2}{2m}\cdot \frac{1}{2} \left( \hat{\psi}^\dagger(x) \hat{\vec{D}}^2_e(x) \hat{\psi}(x)
+ \left( \hat{\vec{D}}^2_e(x) \hat{\psi}(x) \right)^\dagger \cdot \hat{\psi}(x) \right). \label{eq:op_ked}
\end{eqnarray}
This definition is motivated by the relativistic energy dispersion relation $E = \sqrt{(pc)^2 + (mc^2)^2} \approx mc^2 + \frac{p^2}{2m}$ where $E$ and $p$ are the energy and momentum respectively.
We take the kinetic energy part $\frac{p^2}{2m}$ (discard the constant mass term), replace $p_k$ by $i \hbar \hat{D}_{ek}$ as a usual quantization rule under the existence of the electromagnetic field, and construct a field operator by sandwiching between $\hat{\psi}^\dagger(x)$ and $\hat{\psi}(x)$. The field operator is made to be Hermitian by adding the Hermitian conjugate and divided by two, which leads to Eq.~\eqref{eq:op_ked}.

After applying the primary RQED approximation to Eq.~\eqref{eq:op_ked}, we
take the expectation value with respect to the stationary state of the electrostatic Hamiltonian to obtain the kinetic energy density as
\begin{eqnarray}
n_{T_{e}}(\vec{r}) = - \frac{\hbar^2}{4m}
\sum_{i} \nu_i \left[ \psi_{i}^{*}(\vec{r}) \Delta \psi_{i}(\vec{r}) + 
\Delta \psi_{i}^{*}(\vec{r}) \cdot \psi_{i}(\vec{r}) \right].   \label{eq:ked} 
\end{eqnarray}
This definition of the kinetic energy density is not positive-definite so that the whole space is divided into three types of region \cite{Tachibana2001}: (i) the electronic drop region $R_D = \left\{\,\vec{r}\,|\,n_{T_e}(\vec{r}) > 0 \right\}$, where the classically allowed motion of electron is guaranteed and the electron density is amply accumulated, (ii) the electronic atmosphere region $R_A = \left\{\,\vec{r}\,|\,n_{T_e}(\vec{r}) < 0 \right\}$, where the motion of electron is classically forbidden and the electron density is dried up, and (iii) the electronic interface $S = \left\{\,\vec{r}\,|\,n_{T_e}(\vec{r}) = 0 \right\}$, which corresponds to a turning point and is the boundary between $R_D$ and $R_A$.
The outermost $S$ is proposed to give a clear image of the intrinsic shape of atoms and molecules.
These regions, in particular the negative kinetic energy region, can be understood in association with the context of the Wentzel-Kramers-Brillouin (WKB) approximation of the one-dimensional Schr\"{o}dinger equation under the potential energy $V(x)$ \cite{CondonShortley}. 
There, the approximate solution is oscillatory in the positive kinetic energy region
($E-V(x)>0$) and it is exponentially decaying in the negative kinetic energy region ($E-V(x)<0$).

We note that ambiguity regarding the definition of the kinetic energy density has been discussed in the literature \cite{Ayers2002,GarciaAldea2007,Anderson2010}. The kinetic energy density is sometimes defined as a positive definite quantity. 
 We advocate the use of non-positive-definite definition \eqref{eq:ked} as it comes from a field theoretic construction as described above.
 Also, our definition is phenomenologically more useful in a sense that the shape of atoms and molecules can be defined by the zero isosurface of  a non-positive-definite quantity. 
Although it is frequently defined by the isosurface of the electron density, the values of isosurface like 0.001 a.u. and 0.002 a.u. are proposed \cite{Bader1987,Bader,Popelier}, and there is some arbitrariness. 
Since our approach uses the zero isosurface, there is no such arbitrariness. 

In the end of this section, we note on the general merit of our method of using the field theoretic quantities. One of the most important merits is that our method can be applied to not only molecular (finite) systems but also to crystal (periodic) systems. Namely, we can compute  the field theoretic quantities, such as the electronic stress tensor density and kinetic energy density, for both types of systems. 
For the periodic system, the natural orbitals in Eqs.~\eqref{eq:stress}, \eqref{eq:tension} and \eqref{eq:ked} are replaced by the Bloch orbitals \cite{Tachibana2014a}.
Conventionally, the former is analyzed by the molecular orbital theory and the latter by the band theory, but using our method, these systems can be treated by the same field theoretic quantities and we can obtain a unified viewpoint.

%%%%%%%%%%%%%%%%%%%%%%%%%%%%%%%%%%%%%%%%%%%
\section{Results and Discussions} \label{sec:results}
%%%%%%%%%%%%%%%%%%%%%%%%%%%%%%%%%%%%%%%%%%%

%%%%%%%%%%%%%%%%%%%%%%%%%%%%%%%%%%%%%%%%%%%
\subsection{Data set and computational details} \label{sec:data}
%%%%%%%%%%%%%%%%%%%%%%%%%%%%%%%%%%%%%%%%%%%

As for ${\rm Li_3PO_4}$ and ${\rm Li_3PS_4}$, we use the cluster models, respectively, ${\rm (Li_3PO_4)_4}$ and ${\rm (Li_3PS_4)_4}$, shown in Fig.~\ref{fig:lisicon}. They are assumed to be electrically neutral. 
Their geometries are taken from the crystal structures of $\beta$ and $\gamma$-phases of these materials. 
We adopt the lattice constants from the experimental values in Ref.~\cite{Keffer1967} for  $\beta$-${\rm Li_3PO_4}$, Ref.~\cite{Yakubovich1997} for $\gamma$-${\rm Li_3PO_4}$, and Ref.~\cite{Homma2011} for $\beta$ and $\gamma$-${\rm Li_3PS_4}$.
As for $\gamma$-${\rm Li_3PO_4}$ and $\beta$-${\rm Li_3PS_4}$, the models consist of four unit cells, and as for $\beta$-${\rm Li_3PO_4}$ and $\gamma$-${\rm Li_3PS_4}$, two unit cells. 
We note that the observed structure of $\beta$-${\rm Li_3PS_4}$ is a mixture of two structures in which different sites are occupied by Li atom \cite{Homma2011,Mercier1982}:
one with the $4b$ site in the Wyckoff notation occupied, and one with the $4c$ site occupied. They are designated as  $\beta$-${\rm Li_3PS_4}$-$b$ and $\beta$-${\rm Li_3PS_4}$-$c$ respectively in Ref.~\cite{Lepley2013}, and the latter is what we use in this paper.

We also analyze materials which are typically considered to have covalent, metallic and ionic bonds. 
As for them, the geometries of our cluster models are again taken from the crystal structures:
(a) C and Si are modeled by the diamond structure, consists of 42 atoms, 
(b) Li and Na by the body-centered cubic (BCC) structure, consists of 35 atoms, 
and (c) LiF, LiCl, NaF, and NaCl by the sodium-chloride-type structure, consists of 18 metallic atoms and 18 non-metallic atoms. 
The lattice parameters are adopted from the experimental values in Ref.~\cite{Kittel} for (a) and (b), and Ref.~\cite{West} for (c).
These cluster models are depicted in the Fig.~S1 in Supporting Information.

The electronic structures for these models are obtained by the density functional theory (DFT) method using the Lee-Yang-Parr gradient-corrected functional \cite{Lee1998,Miehlich1989} with Becke's three hybrid parameters \cite{Becke1993} (B3LYP). We use the 6-31++G(d,p) basis set \cite{Ditchfield71,Hehre72,Hariharan73,Francl1982,Frisch84} for ${\rm Li_3PO_4}$ and ${\rm Li_3PS_4}$, and the 6-311++G(d,p) basis set \cite{Raghavachari80b,McLean1980,Frisch84} for the other materials. The spin multiplicities are taken to be singlet, except for Li and Na which are taken to be doublet.

We use the Gaussian 09 \cite{Gaussian09} for the computation of the electronic structures.
To compute the aforementioned quantities 
such as Eqs.~\eqref{eq:stress}, \eqref{eq:tension} and \eqref{eq:ked} from the electronic structure data,
we use the QEDynamics package \cite{QEDynamics} developed in our group.

%%%%%%%%%%%%%%%%%%%%%%%%%%%%%%%%%%%%%%%%%%%
%\subsection{Differential eigenvalues of electronic stress tensor density at Lagrange point} \label{sec:3.2}
\subsection{Eigenvalues of the electronic stress tensor density at the Lagrange point} \label{sec:eig}
%%%%%%%%%%%%%%%%%%%%%%%%%%%%%%%%%%%%%%%%%%%
In this section, as is customarily done in our previous works \cite{Szarek2007, Szarek2008, Szarek2009, Ichikawa2011a, Ichikawa2011b, Ichikawa2011c,Ichikawa2012,Nozaki2015}, we analyze the chemical bonds of ${\rm Li_3PO_4}$ and ${\rm Li_3PS_4}$ by searching for a Lagrange point between atomic pairs and computing the eigenvalues of the electronic stress tensor density there.
In order to avoid too many data points, we limit our search to four pairs around a Li atom and four pairs around a P atom in each material. 
These pairs are shown in Fig.~\ref{fig:lisicon} as ball-and-stick model and labelled with numbers. 
The results are summarized in Table \ref{tab:bond_lisicon} and Figs.~\ref{fig:dist_eig} and \ref{fig:diff_eig}.
Fig.~\ref{fig:dist_eig} shows the relation between bond distance and the largest eigenvalue, and Fig.~\ref{fig:diff_eig} is a scatter plot of the differential eigenvalues,
$\tau^{S33}_{e} -  \tau^{S22}_{e}$ and $\tau^{S22}_{e} -  \tau^{S11}_{e}$.
We here note on the effect of spin-multiplicity on our results. As the valency of atoms on the edge of the cluster is not satisfied, the singlet states we adopt are not the ground states. We repeat our analysis for the higher spin states which have lower energy than the singlet states, and the results are shown in Table S1 and Fig.~S2. We see that the difference does not affect the qualitative arguments below.
In Figs.~\ref{fig:dist_eig} and \ref{fig:diff_eig}, in addition to the data of ${\rm Li_3PO_4}$ and ${\rm Li_3PS_4}$, we also plot those for alkali metal clusters, hydrocarbons, and bonds between Ge, Sb and Te atoms in GeSbTe (GST) molecules taken from previous studies \cite{Ichikawa2012,Nozaki2015}. 
We furthermore include data for C, Si, Li, Na, and ionic compounds (LiF, LiCl, NaF, NaCl) described in the previous subsection. 
The location of the atomic pair used for our computation is shown in Fig.~S1.
Their data are plotted in Figs.~\ref{fig:dist_eig} and \ref{fig:diff_eig}, and summarized in Table \ref{tab:bond_others}.

Let us first review our past studies using Figs.~\ref{fig:dist_eig} and \ref{fig:diff_eig}.
We see in Fig.~\ref{fig:dist_eig} that most of the bonds in hydrocarbon molecules have positive largest eigenvalues (tensile stress), which is consistent with the spindle structure and covalency, as mentioned in Sec.~\ref{sec:intro}. Although we see some bonds with negative eigenvalues among the hydrocarbon molecules, the eigenvalue can be negative for a covalent bond typically in a very short bond such as a triple bond in C$_2$H$_2$ \cite{Tachibana2005,Szarek2008,Ichikawa2011b}.  
This is because the region between the C atoms of C$_2$H$_2$ is overwhelmed by the atomic compressive stress around the C nuclei as they are so close. The atomic compressive stress in turn is attributed to marginal stability around atoms \cite{Tachibana2005}.
The electronic stress tensor density of atoms has been systematically investigated recently in Ref.~\cite{Nozaki2015b}, to which we refer for the details. 
As for the alkali metal clusters, their largest eigenvalues are negative (compressive stress), which characterize one aspect of metallicity in terms of the electronic stress tensor density \cite{Szarek2007,Ichikawa2012,Tachibana2013}. 
However, we have to see the differential eigenvalues to tell whether the negativity of the largest eigenvalue originates from the short bond (atomic stability) or metallicity. In fact, in Fig.~\ref{fig:diff_eig}, we see that the Li and Na clusters have very small $\tau^{S 33}_{e}-\tau^{S 22}_{e}$ and $\tau^{S 22}_{e}-\tau^{S 11}_{e}$ which are much smaller than $\tau^{S 33}_{e}-\tau^{S 22}_{e}$ of hydrocarbons.
Namely, as mentioned in Sec.~\ref{sec:intro}, the three eigenvalues of the Li and Na clusters have almost same values
while the hydrocarbon molecules have the largest eigenvalue much larger than the second largest eigenvalue, which has similar value to the smallest eigenvalue. The former degeneracy pattern suggests that the bonds are not directional reflecting the metallicity, and the latter implies the clear directionality of the bonds as expected for covalency \cite{Ichikawa2012}.
We note that this has been found in Ref.~\cite{Ichikawa2012} using the electronic structure computed by the coupled-cluster single-double (CCSD) method.
The result in this paper using the DFT method with B3LYP is qualitatively similar to the one using the CCSD method. The comparison of the differential eigenvalues for the Li clusters is shown in Fig.~S3.
As for the GST molecules, the largest eigenvalues can be both positive and negative and the degeneracy pattern is intermediate between the alkali metal and hydrocarbons. This is consistent with the usual classification of Ge, Sb, and Te as metalloids \cite{Nozaki2015}. 

We now move on to examine the bonds in ${\rm Li_3PO_4}$ and ${\rm Li_3PS_4}$.
The P-O bonds have the largest eigenvalues which are negative. As the differential eigenvalues show the degeneracy pattern similar to the hydrocarbons, these negative eigenvalues are likely to be interpreted in the same way as C$_2$H$_2$. We wish to investigate more on this issue in the next subsection by looking at the eigenvalue map in the region including the P and O nuclei. 
The largest eigenvalues of the Li-O bonds are positive, but they are small compared to the typical hydrocarbons. The degeneracy pattern is in between the hydrocarbons and alkali metal clusters.
Therefore, we may say that, in view of the eigenvalue patterns at the Lagrange point, the Li-O bonds look like the GST bonds. The resemblance is stronger for the Li-S bonds: their largest eigenvalues can be positive and negative whose absolute values are close to those of the GST bonds, and the degeneracy pattern is also similar to the GST bonds. However, since the Li-O and Li-S are considered to be ionic while the GST bonds are not,
they should be discriminated by some other means. This will be discussed in the following subsection.
As for the P-S bonds, the largest eigenvalues are positive, and the absolute values distribute in a similar range to those of the P-O bonds. The differential eigenvalues are also relatively close to the hydrocarbons and P-O bonds. On the closer look, the P-S bonds are slightly more degenerate than the P-O bonds, but less degenerate than the ionic bonds or bonds among metalloids. 

In the end of this section, we comment on the data for C, Si, Li, Na, and ionic compounds (LiF, LiCl, NaF, NaCl).
The data for C are obviously similar to those for the hydrocarbons, and 
the data for Si, which is usually classified as metalloids, are consistently similar to those for the GST bonds.
The data for Li and Na are of course fall in the range of the data for the alkali metals.  
As for the ionic compounds, their data are similar to those for the Li-O bonds and Li-S bonds, and to those for the GST bonds.

%%%%%%%%%%%%%%%%%%%%%%%%%%%%%%%%%%%%%%%%%%%
\subsection{Electronic stress tensor density and kinetic energy density} \label{sec:strked}
%%%%%%%%%%%%%%%%%%%%%%%%%%%%%%%%%%%%%%%%%%%
In this section, we examine the electronic stress tensor density not only at the Lagrange point and but also in the region encompassing atomic pairs. 
We also investigate the electronic kinetic energy density, in particular its zero surface, the electronic interface denoted by $S$. 

In Fig.~\ref{fig:bLi3PO4}, we show the electronic stress tensor density and kinetic energy density for bonds in $\beta$-${\rm Li_3PO_4}$.
The panels (a) and (b) include the Li-O bonds, and the panels (c) and (d) the P-O bonds. 
The largest eigenvalue of the electronic stress tensor density and corresponding eigenvector are plotted in the left column, and the electronic kinetic energy density is plotted in the right column. 
We plot similarly for $\beta$-${\rm Li_3PS_4}$ in Fig.~\ref{fig:bLi3PS4}.
As for $\gamma$-${\rm Li_3PO_4}$ and $\gamma$-${\rm Li_3PS_4}$, since they are not significantly different from the $\beta$-phase counterparts for our discussion, we show the figures in Supporting Information Figs.~S5 and S7 (we also plot higher resolution version of Figs.~\ref{fig:bLi3PO4} and \ref{fig:bLi3PS4} in Figs.~S4 and S6 respectively).
We note that, when only the sign of the kinetic energy density matters, we can tell it by recognizing the region in the neighborhood of the nucleus should have positive kinetic energy density, that is to say $R_D$. Going outward, $R_A$ and $R_D$ appear alternately bounded by $S$ and we should have $R_A$ at infinity. In this paper, we show the kinetic energy density map for clarity. 
In Fig.~S8, we show a comparison between the DFT method with B3LYP and the CCSD method for a Li$_4$ cluster regarding the electronic stress tensor density and kinetic energy density. 
At least for this cluster, these two methods give almost identical results.

We first point out that the P-S bonds in the panels (c) and (d) of Fig.~\ref{fig:bLi3PS4} exhibit typical spindle structures \cite{Tachibana2004} mentioned in Sec.~\ref{sec:intro}. There is a region between the P and S atoms with the positive eigenvalues and corresponding eigenvectors are in the direction such as to connect the atomic pair. The Lagrange point is located in the spindle structure consistently with what is found in the previous subsection. 
The P-O bonds in the panels (c) and (d) of Fig.~\ref{fig:bLi3PO4} exhibit similar but slightly different pattern. It is distorted away from P toward O, and the Lagrange point is located in the negative eigenvalue region around the P atom. 
Therefore, we may consider the reason why the P-O bonds have negative eigenvalue at the Lagrange point is same as that of the C-C bond in C$_2$H$_2$ in a sense that it is overwhelmed by the atomic compressive stress around the nuclei. However, in the case of P-O bonds, it is not due to a very short bond and rather caused by the size difference between P and O. 

We next study the Li-O and Li-S bonds which are considered to be ionic. 
The pattern of the eigenvectors is connecting the atomic nuclei for all these bonds, but that of eigenvalue looks somewhat irregular. 
The conspicuous difference from the P-O and P-S bonds is found in the pattern of $S$.
This can be easily recognized in the right columns of Figs.~\ref{fig:bLi3PO4} and \ref{fig:bLi3PS4}. 
In fact, we see in the panels (c) and (d) of Fig.~\ref{fig:bLi3PO4} that a connected $S$ encloses the P-O bonds whereas $S$'s around Li and O are disconnected as shown in the panels (a) and (b). 
In other words, the $R_D$'s forming around Li and O are separated by $R_A$, while P and O form a connected $R_D$ (strictly speaking, there is a ring-like $R_A$ surrounding the P nucleus, but this is just a partition of the P's core and valence regions. The P's valence region and O are connected by $R_D$). 
We can see in Fig.~\ref{fig:bLi3PS4} that the same pattern is true for the P-S bonds and Li-S bonds.
We note that, although we cannot recognize from these two-dimensional figures, the $R_D$ of Li looks like a ball in three-dimensional space and it is not connected to the $R_D$ formed from P and O. We have checked this by drawing $S$ in three-dimensional space.

Some comments on the $R_D$ of Li may be in order. 
One may wonder why it has the positive largest eigenvalue despite the atomic compressive stress aforementioned in the previous subsection.  
As discussed in detail in Ref.~\cite{Nozaki2015b}, this is an exceptional case due to the Lewis electron paring of $(1s)^2$ in the Li core.  
In fact, a He atom and the core region of a Be atom have the same eigenvalue pattern. 
As the nuclear charge increases, the tensile stress caused by the electron paring is immersed under the atomic compressive stress, and we do not see the positive eigenvalue region at the center of, for instance, O and P.
Therefore, we can interpret the $R_D$ of Li filled with positive largest eigenvalue seen in ${\rm Li_3PO_4}$ and ${\rm Li_3PS_4}$ as the manifestation of the fact that Li exists as a Li cation, which should be the case for the ionic bonding. 

The morphology of $S$ pointed out above in the ionic bond is not surprising because, in the ionic bond, the cation and anion are bonded by their electrostatic interaction which does not require the bonding region with the electron density is amply accumulated.
In our terms, the $R_D$'s which belong to the cation and anion may be separated by $R_A$.
This is contrary to the covalent or metallic bonding where the nuclei are bonded by some amount of electron density in between. Namely, the atoms involved in these bonds are connected by $R_D$ and are enclosed by a single $S$.

Thus, we now have a way to discriminate the ionic bonds and the bonds between metalloid atoms, which are indistinguishable by the eigenvalue pattern at the Lagrange point as argued in the previous subsection. 
That is, we can do this by looking at $S$ involved in the bond: it is disconnected in the former case and connected in the latter case. 
In fact, as we have computed in Ref.~\cite{Nozaki2015} (see also its supporting information), all the $S$'s of the GST molecules are connected.

Lastly, we wish to confirm this idea by looking at the other cluster models whose bonds are typically covalent, metallic, and ionic. 
In Fig.~\ref{fig:CSiLiNa}, we show the electronic stress tensor density and kinetic energy density for bonds in C, Si, Li, and Na clusters, and we show those for ionic clusters of LiF, LiCl, NaF, and NaCl, in Fig.~\ref{fig:ionic} (higher resolution version is found in Supporting Information Figs.~S9 and S10). 
As seen in the right columns of Fig.~\ref{fig:CSiLiNa}, the atoms are connected by $R_D$ for the covalent and metallic bonds.
Note that, as mentioned above in the case of P in ${\rm Li_3PO_4}$, the ring-like $R_A$'s (they are thin-shell-like regions in three-dimensional space) surrounding the atomic nuclei are just partitions of the core and valence regions \cite{Nozaki2015b}. The valence regions of those clusters are connected by $R_D$.
On the contrary, showing in Fig.~\ref{fig:ionic}, there is $R_A$ which separates $R_D$'s of cations and anions for all the ionic clusters, supporting what we have inferred from the ionic bonds in ${\rm Li_3PO_4}$ and ${\rm Li_3PS_4}$.
As noted above, the $R_D$ of Li has positive eigenvalue as expected for the Li core (consistently, similar core regions are found for the Li cluster as in Fig.~\ref{fig:CSiLiNa} (c) ). 
In addition, one may notice that the outermost region of $R_D$, in particular on $S$, of Na and F also has positive eigenvalue.
This pattern is quite close to that of a Ne atom, which is caused by the electron parings in $2p_x^2$,  $2p_y^2$, and $2p_z^2$ \cite{Nozaki2015b}, and it is reasonable to find such a pattern in a Na cation and a F anion in their ionic compounds. 

Let us conclude this section by intuitive explanations of ionic character of chemical bonding using the electronic kinetic energy density. 
Our kinetic energy density, which is defined to be non-positive-definite, can distinguish the region
with high electron density (positive region) and low electron density (negative region).
Intuitively, ions are spherical and they can be regarded as composed of two parts: 
a central core part in which most of the electron density is concentrated and an outer part which 
contains very little electron density.
Therefore, in terms of the kinetic energy density, an ion exhibits almost spherical surface on which the kinetic energy density is zero, 
and positive inside. Namely, the ion core is represented by a ball-like region with positive kinetic energy density.
Then, ionic clusters or crystals are described by arrays of such ball-like regions.
Between these balls, there is a negative kinetic energy region and this is reasonable 
because they, representing anions and cations, are bonded by the electrostatic interaction and accumulation of the electron density is not required. 
Such pattern is clearly seen in Fig. 7, on the right panels. 

%%%%%%%%%%%%%%%%%%%%%%%%%%%%%%%%%%%%%%%%%%%
\section{Conclusion} \label{sec:conclusion}
%%%%%%%%%%%%%%%%%%%%%%%%%%%%%%%%%%%%%%%%%%%
We have analyzed the electronic structure of $\beta$ and $\gamma$-phases of ${\rm Li_3PO_4}$ and ${\rm Li_3PS_4}$, which are interested in as lithium ionic conductors, using local quantities derived by quantum field theoretic consideration, the electronic stress tensor density and kinetic energy density.

We have first found that, as long as we examine the pattern of the eigenvalues of the electronic stress tensor density at the Lagrange point, we cannot distinguish between the ionic bonds and bonds among metalloid atoms. In particular, both of them exhibit  the pattern of the differential eigenvalues which is intermediate between covalent and metallic bonding. 

We have then proposed that they can be distinguished by looking at the morphology of the electronic interface, $S$, the zero surface of the electronic kinetic energy density.  
In the covalent and metallic bonds, including those between metalloid atoms, the atoms participating in the bonds are enclosed by a single $S$. In other words, the electronic drop region $R_D$, where the electronic kinetic energy density is positive, of the atoms is connected.  
On the other hand, in the ionic bonds, the cations and anions are bounded by each own $S$. Namely, their $R_D$'s are disconnected and there exists the electronic atmosphere region $R_A$, where the electronic kinetic energy density is negative, in between them.
We have argued that this characterization is reasonable for the ionic bonds, because
the cation and anion are bonded by their electrostatic interaction which does not require the bonding region with the accumulated electron density.

In this paper, we have shown that the ionicity of chemical bonding can be characterized by the electronic stress tensor density and kinetic energy density, in addition to the covalency and metallicity which had been discussed in our previous works. 
The findings regarding the ionicity here have much room for further research, for example, definition of the ionic radius, and 
quantification of partial ionic character of covalent bonds, which will be pursued in our future works.
The results in this paper are based on the computation at the equilibrium positions.  
We would have much more insight by analyzing the electronic stress tensor density and kinetic energy density as the internuclear distance changes for the ionic bonds. 
This will be studied in future by using the two lowest states of the alkali halides molecules along the potential energy curves computed with the configuration interaction method. 

In the end, we stress that our studies are based on the quantities defined at each point in space which originate from the quantum field theoretic consideration, not from the electron density. We believe that our method will not only lead us to more fundamental understanding of the chemical properties of the lithium ionic conductors but also serve as useful characterization tools for the materials. 

\noindent 
%%%%%%%%%%%%%%%%%%%%%%%%%%%%%%%%%%%%
\section*{Acknowledgment}
%%%%%%%%%%%%%%%%%%%%%%%%%%%%%%%%%%%%
Theoretical calculations were partly performed using Research Center for Computational Science, Okazaki, Japan.
This work was supported by JSPS KAKENHI Grant Number 25410012 and 26810004.
This work was partly supported by funding from Samsung R\&D Institute Japan.
H.~N. is supported by the Sasakawa Scientific Research Grant from the Japan Science Society.

\clearpage

%%%%%%%%%%%%%%%% table %%%%%%%%%%%%%%%%

\begin{table}[htpb]
\caption{Data for the bonds in ${\rm Li_3PO_4}$ and ${\rm Li_3PS_4}$.
$r_e$ is the bond distance.
$\tau_e^{S33}$, $\tau_e^{S22}$, and $\tau_e^{S11}$ are three eigenvalues
of the electronic stress tensor density at the Lagrange point.}
\label{tab:bond_lisicon}

\vspace{0.5cm}

%\begin{tabular}{cccccc}
\begin{tabular}{l l | r r r r}

\hline
Compounds & Atomic pair & $r_e$[\AA] & $\tau_e^{S33}$ & $\tau_e^{S22}$ & $\tau_e^{S11}$ \\
\hline
$\beta$-${\rm Li_3PO_4}$
&Li(13)-O(35)	&2.015	&$ 8.41\times 10^{-3}$	&$-1.90\times 10^{-2}$	&$-2.00\times 10^{-2}$\\
&Li(13)-O(45)	&2.015	&$ 8.38\times 10^{-3}$	&$-1.90\times 10^{-2}$	&$-1.99\times 10^{-2}$\\
&Li(13)-O(52)	&2.014	&$ 8.68\times 10^{-3}$	&$-1.95\times 10^{-2}$	&$-2.05\times 10^{-2}$\\
&Li(13)-O(56)	&1.958	&$ 8.59\times 10^{-3}$	&$-2.11\times 10^{-2}$	&$-2.24\times 10^{-2}$\\
&P(33)-O(39)	&1.545	&$-2.87\times 10^{-3}$	&$-2.54\times 10^{-1}$	&$-2.55\times 10^{-1}$\\
&P(33)-O(43)	&1.545	&$-2.87\times 10^{-3}$	&$-2.54\times 10^{-1}$	&$-2.55\times 10^{-1}$\\
&P(33)-O(51)	&1.550	&$-2.67\times 10^{-3}$	&$-2.49\times 10^{-1}$	&$-2.50\times 10^{-1}$\\
&P(33)-O(56)	&1.539	&$-7.09\times 10^{-3}$	&$-2.58\times 10^{-1}$	&$-2.58\times 10^{-1}$\\

\hline
$\gamma$-${\rm Li_3PO_4}$
&Li(5)-O(44)	&2.009	&$ 8.38\times 10^{-3}$	&$-1.93\times 10^{-2}$	&$-2.05\times 10^{-2}$\\
&Li(5)-O(47)	&1.942	&$ 7.50\times 10^{-3}$	&$-1.99\times 10^{-2}$	&$-2.08\times 10^{-2}$\\
&Li(5)-O(53)	&1.945	&$ 8.52\times 10^{-3}$	&$-2.10\times 10^{-2}$	&$-2.19\times 10^{-2}$\\
&Li(5)-O(61)	&1.924	&$ 8.00\times 10^{-3}$	&$-2.18\times 10^{-2}$	&$-2.30\times 10^{-2}$\\
&P(27)-O(35)	&1.550	&$-1.36\times 10^{-3}$	&$-2.53\times 10^{-1}$	&$-2.56\times 10^{-1}$\\
&P(27)-O(41)	&1.550	&$-2.54\times 10^{-3}$	&$-2.49\times 10^{-1}$	&$-2.49\times 10^{-1}$\\
&P(27)-O(53)	&1.542	&$-2.03\times 10^{-3}$	&$-2.61\times 10^{-1}$	&$-2.63\times 10^{-1}$\\
&P(27)-O(63)	&1.545	&$-1.04\times 10^{-2}$	&$-2.40\times 10^{-1}$	&$-2.41\times 10^{-1}$\\

\hline
$\beta$-${\rm Li_3PS_4}$
&Li(9)-S(41)	&2.758	&$ 3.93\times 10^{-4}$	&$-2.34\times 10^{-3}$	&$-2.50\times 10^{-3}$\\
&Li(9)-S(45)	&2.179	&$-2.52\times 10^{-3}$	&$-1.46\times 10^{-2}$	&$-1.58\times 10^{-2}$\\
&Li(9)-S(53)	&2.500	&$ 3.33\times 10^{-4}$	&$-5.59\times 10^{-3}$	&$-6.08\times 10^{-3}$\\
&Li(9)-S(63)	&2.300	&$-1.31\times 10^{-3}$	&$-9.69\times 10^{-3}$	&$-1.04\times 10^{-2}$\\
&P(29)-S(37)	&2.014	&$ 7.55\times 10^{-3}$	&$-8.02\times 10^{-2}$	&$-8.09\times 10^{-2}$\\
&P(29)-S(47)	&2.014	&$ 6.82\times 10^{-3}$	&$-8.05\times 10^{-2}$	&$-8.13\times 10^{-2}$\\
&P(29)-S(53)	&2.025	&$ 6.54\times 10^{-3}$	&$-7.81\times 10^{-2}$	&$-7.86\times 10^{-2}$\\
&P(29)-S(61)	&2.069	&$ 9.50\times 10^{-3}$	&$-7.35\times 10^{-2}$	&$-7.49\times 10^{-2}$\\

\hline
$\gamma$-${\rm Li_3PS_4}$
&Li(2)-S(5)	&2.479	&$-1.46\times 10^{-5}$	&$-4.75\times 10^{-3}$	&$-5.25\times 10^{-3}$\\
&Li(2)-S(16)	&2.554	&$ 3.65\times 10^{-4}$	&$-4.55\times 10^{-3}$	&$-4.86\times 10^{-3}$\\
&Li(2)-S(22)	&2.437	&$ 3.55\times 10^{-4}$	&$-7.01\times 10^{-3}$	&$-7.09\times 10^{-3}$\\
&Li(2)-S(45)	&2.554	&$ 3.82\times 10^{-4}$	&$-4.57\times 10^{-3}$	&$-4.89\times 10^{-3}$\\
&P(44)-S(45)	&2.060	&$ 1.16\times 10^{-2}$	&$-7.14\times 10^{-2}$	&$-7.18\times 10^{-2}$\\
&P(44)-S(47)	&2.020	&$ 7.93\times 10^{-3}$	&$-7.88\times 10^{-2}$	&$-8.01\times 10^{-2}$\\
&P(44)-S(48)	&2.060	&$ 1.16\times 10^{-2}$	&$-7.11\times 10^{-2}$	&$-7.16\times 10^{-2}$\\
&P(44)-S(62)	&1.983	&$ 2.55\times 10^{-3}$	&$-8.59\times 10^{-2}$	&$-8.63\times 10^{-2}$\\

\hline
\end{tabular}
\end{table}

\begin{table}[htpb]
\caption{Data for the bonds in C, Si, Li, Na, and ionic compounds (LiF, LiCl, NaF, NaCl). 
$r_e$ is the bond distance.
$\tau_e^{S33}$, $\tau_e^{S22}$, and $\tau_e^{S11}$ are three eigenvalues
of the electronic stress tensor density at the Lagrange point.}
\label{tab:bond_others}

\vspace{0.5cm}

\begin{tabular}{l | r r r r}

\hline
Compounds & $r_e$[\AA] & $\tau_e^{S33}$ & $\tau_e^{S22}$ & $\tau_e^{S11}$ \\
\hline
C
&1.545	&$ 7.32\times 10^{-2}$		&$-1.56\times 10^{-1}$	&$-1.57\times 10^{-1}$\\
Si
&2.351	&$ 1.15\times 10^{-3}$		&$-2.50\times 10^{-2}$	&$-2.51\times 10^{-2}$\\
\hline
Li
&3.023 	&$-2.75\times 10^{-4}$		&$-7.26\times 10^{-4}$	&$-7.26\times 10^{-4}$\\
Na
&3.659 	&$-1.55\times 10^{-4}$		&$-2.55\times 10^{-4}$	&$-2.57\times 10^{-4}$\\
\hline
LiF
&2.014 	&$ 7.55\times 10^{-3}$		&$-2.25\times 10^{-2}$	&$-2.25\times 10^{-2}$\\
LiCl
&2.570 	&$ 8.39\times 10^{-5}$		&$-4.32\times 10^{-3}$	&$-4.32\times 10^{-3}$\\
NaF
&2.320 	&$ 6.30\times 10^{-3}$		&$-2.10\times 10^{-2}$	&$-2.10\times 10^{-2}$\\
NaCl
&2.820 	&$-7.56\times 10^{-4}$		&$-3.78\times 10^{-3}$	&$-3.78\times 10^{-3}$\\
\hline
\end{tabular}
\end{table}

\clearpage

%%%%%%%%%%%%%%%% figure %%%%%%%%%%%%%%%%

\ifFIG	

\begin{figure}
\begin{center}
\includegraphics[width=14cm]{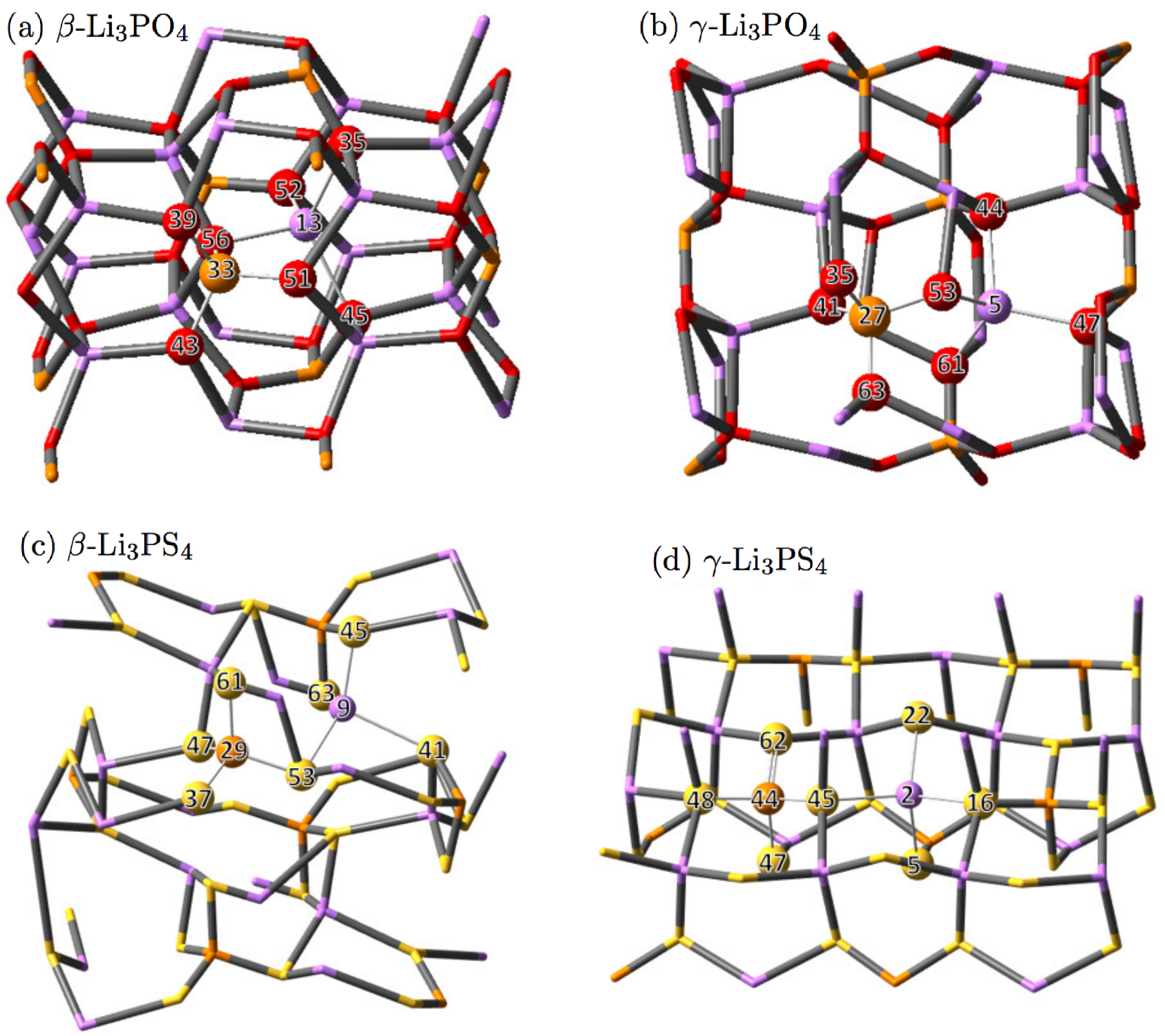}
\caption{
The structures for (a) $\beta$-${\rm Li_3PO_4}$, (b) $\gamma$-${\rm Li_3PO_4}$, (c) $\beta$-${\rm Li_3PS_4}$, and (d) $\gamma$-${\rm Li_3PS_4}$,
which are adopted in this paper. Li is colored purple, O is red, P is orange, and S is yellow. 
Atomic pairs for which we search the Lagrange point and calculate the electronic stress tensor density are shown with ball-and-stick model,
and others are shown with tube model.}
\label{fig:lisicon}
\end{center}
\end{figure}

\begin{figure}
\includegraphics[width=14cm]{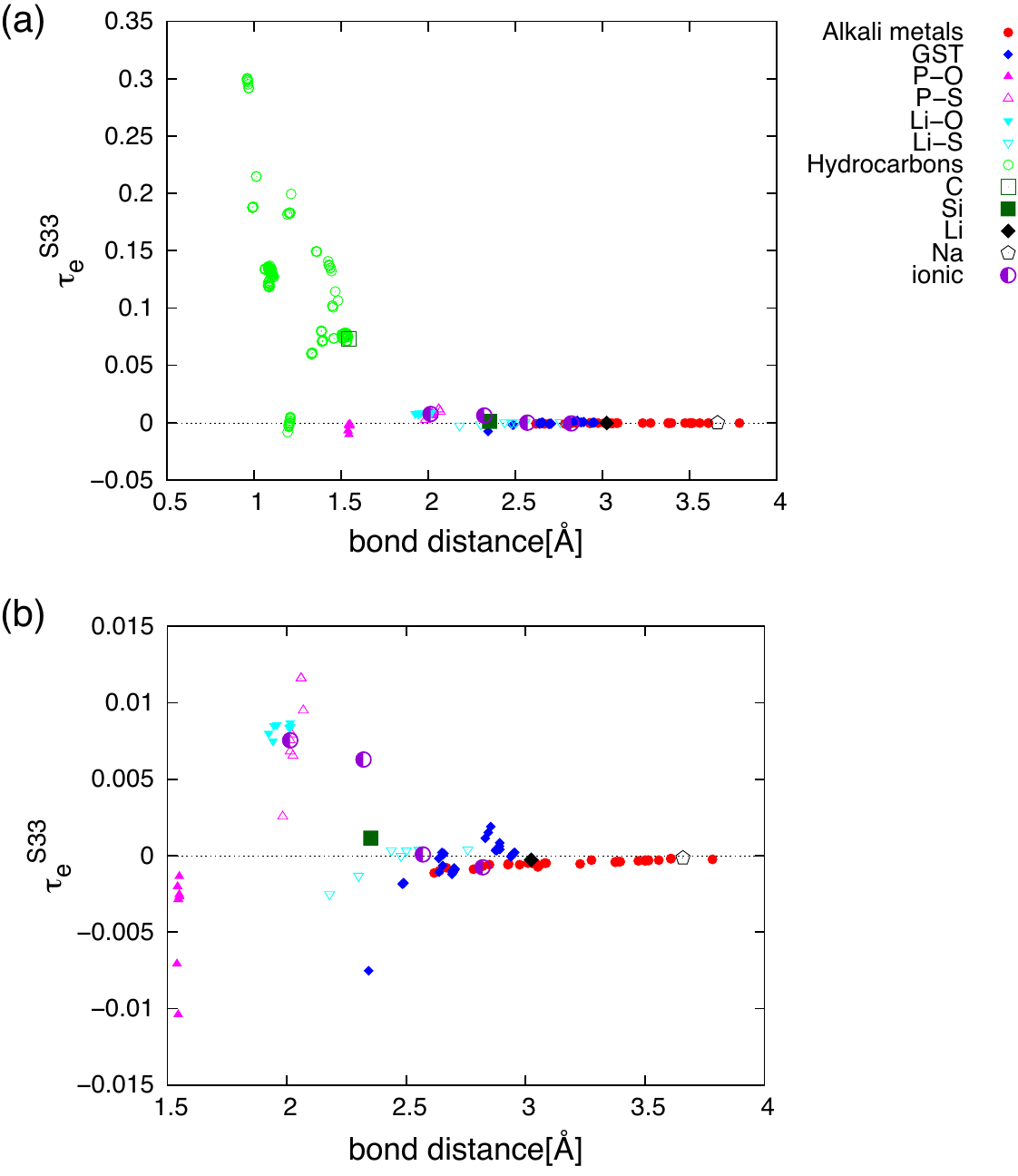}
\caption{The relation between bond distance and the largest eigenvalue of the electronic stress tensor at the Lagrange point. 
The panel (b) is the enlarged view of a part of panel (a).
}
\label{fig:dist_eig}
\end{figure}

\begin{figure}
\includegraphics[width=14cm]{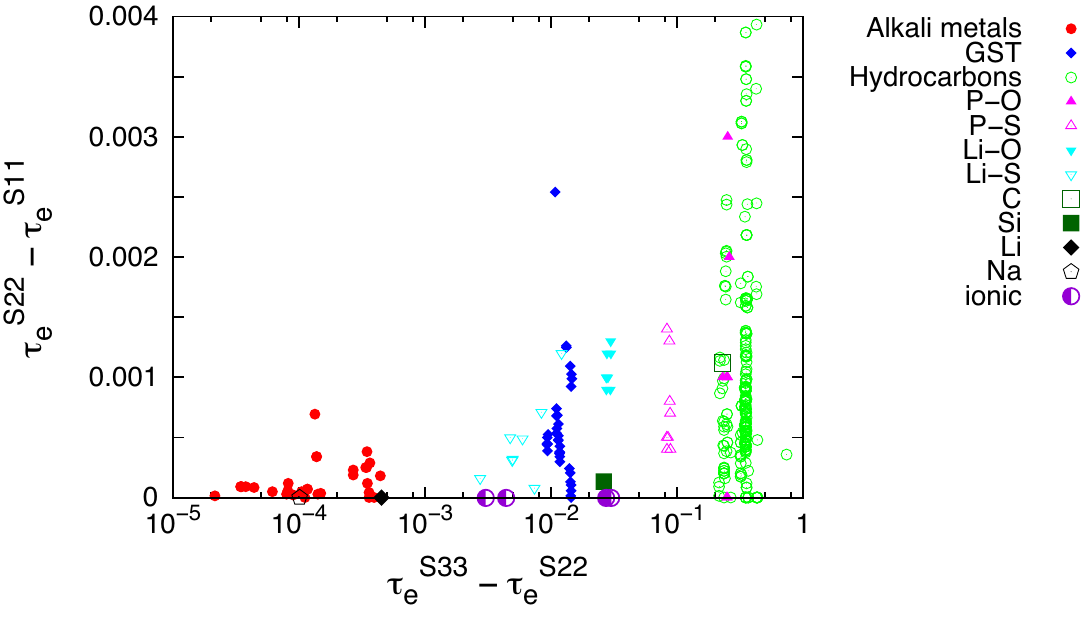}
\caption{Differential eigenvalues of the electronic stress tensor at the Lagrange point. 
}
\label{fig:diff_eig}
\end{figure}

\begin{figure}
\includegraphics[width=12cm]{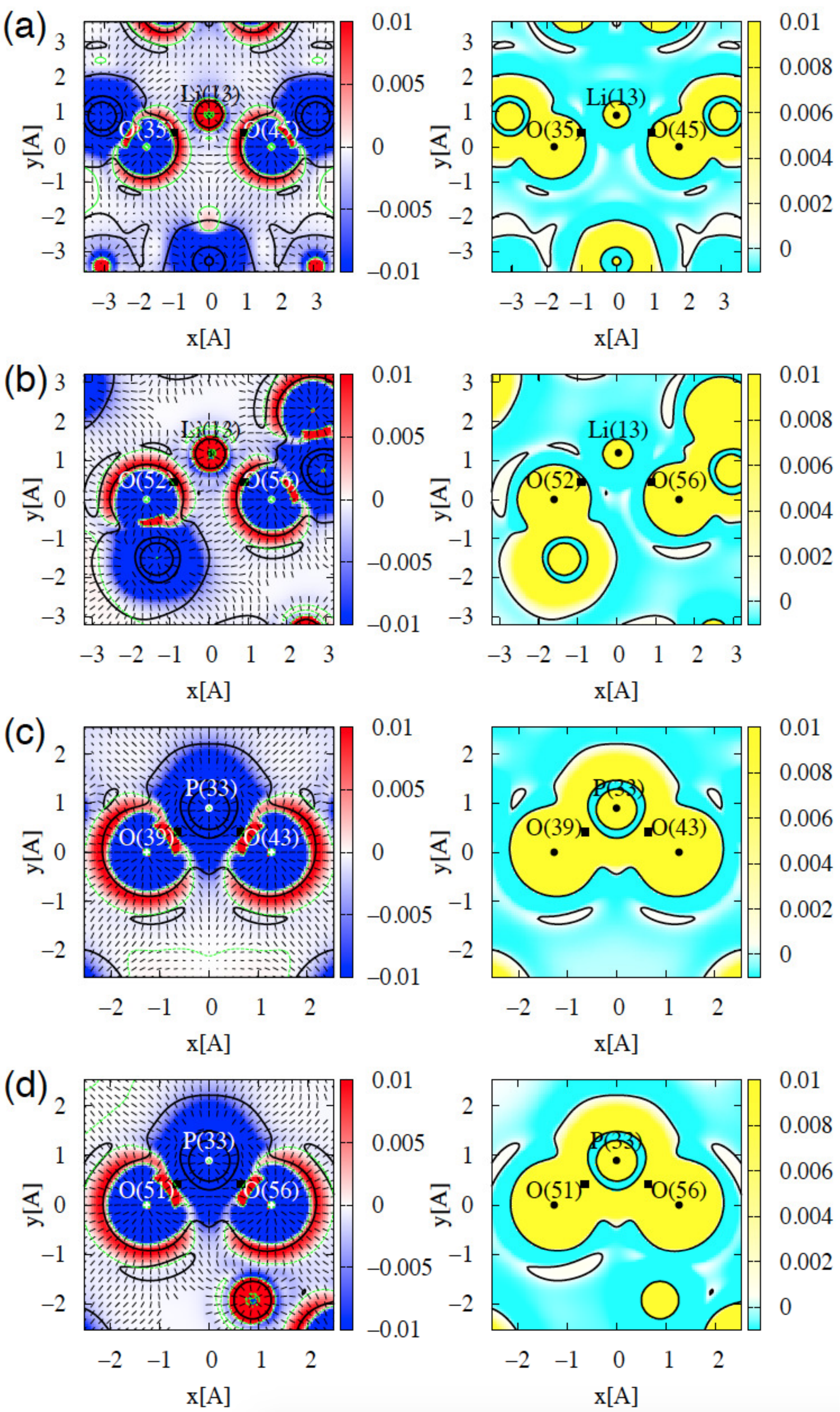}
\caption{The electronic stress tensor density (left column) and the kinetic energy density (right column) for $\beta$-${\rm Li_3PO_4}$.
The panels in the left column show the largest eigenvalue of the electronic stress tensor density (color map) and corresponding eigenvector (black rod) on the plane including the labeled atoms. The right column shows the kinetic energy density by the color map. The black solid lines are the zero surfaces of the kinetic energy density (electronic interface), the green dashed lines are the zero surfaces of the largest eigenvalue, the filled circles show the atomic positions, and the black squares are the Lagrange points.
}
\label{fig:bLi3PO4}
\end{figure}

\begin{figure}
\includegraphics[width=12cm]{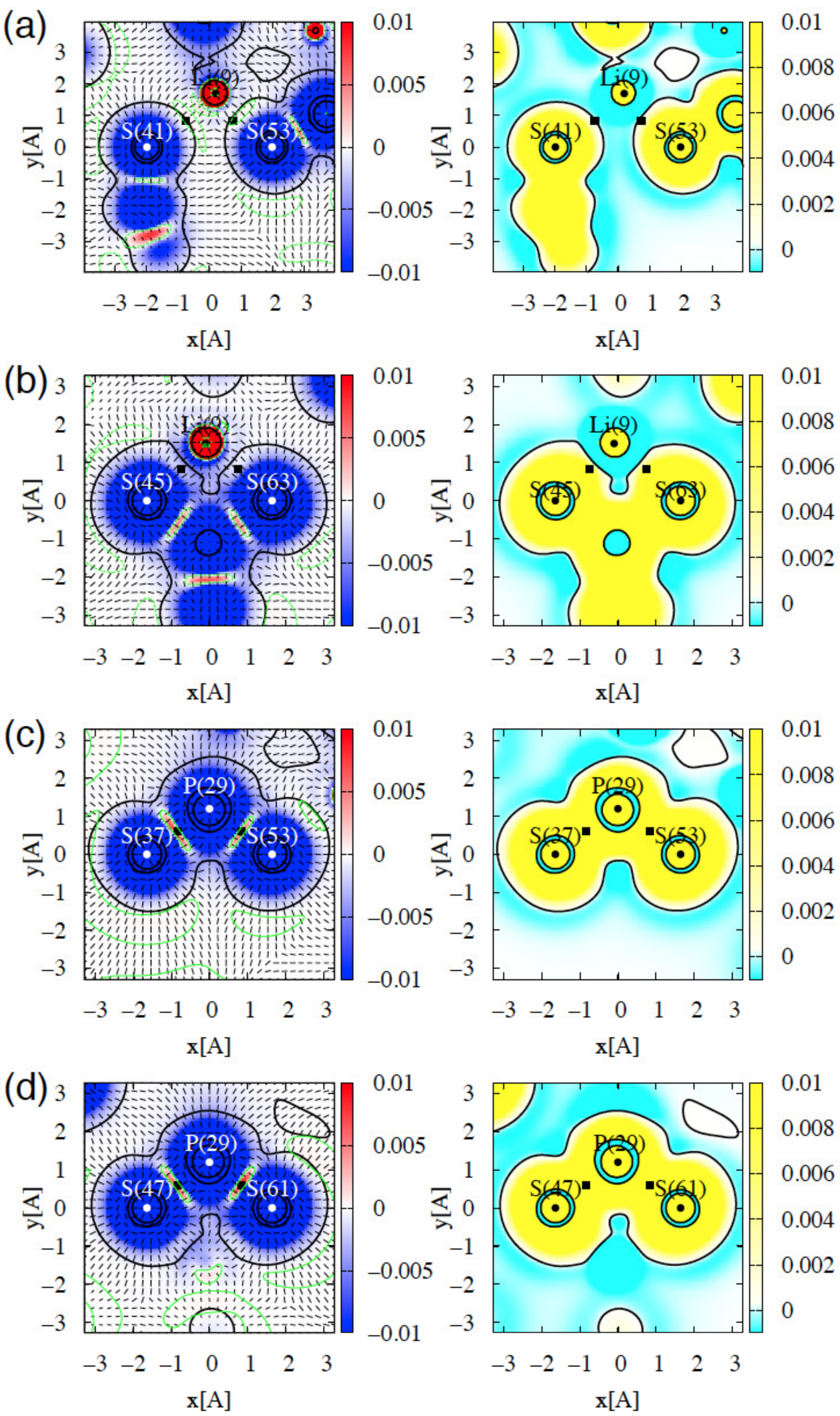}
\caption{The electronic stress tensor density (left column) and the kinetic energy density (right column) for $\beta$-${\rm Li_3PS_4}$ are shown in the same manner as Fig.~\ref{fig:bLi3PO4}.
}
\label{fig:bLi3PS4}
\end{figure}

\begin{figure}
\includegraphics[width=12cm]{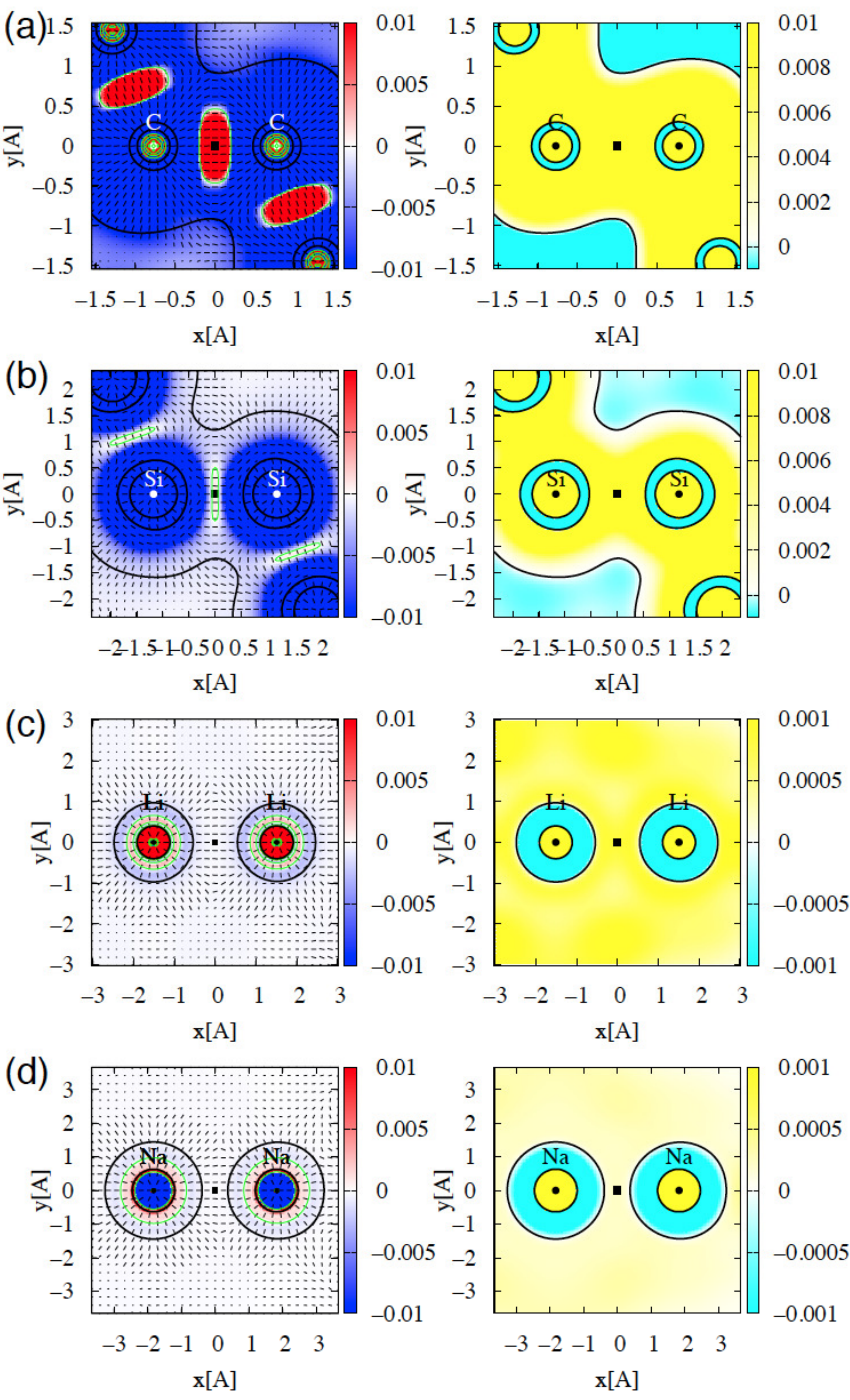}
\caption{The electronic stress tensor density (left column) and the kinetic energy density (right column) for (a) C, (b) Si, (c) Li, and (d) Na are shown in the same manner as Fig.~\ref{fig:bLi3PO4}.
}
\label{fig:CSiLiNa}
\end{figure}

\begin{figure}
\includegraphics[width=12cm]{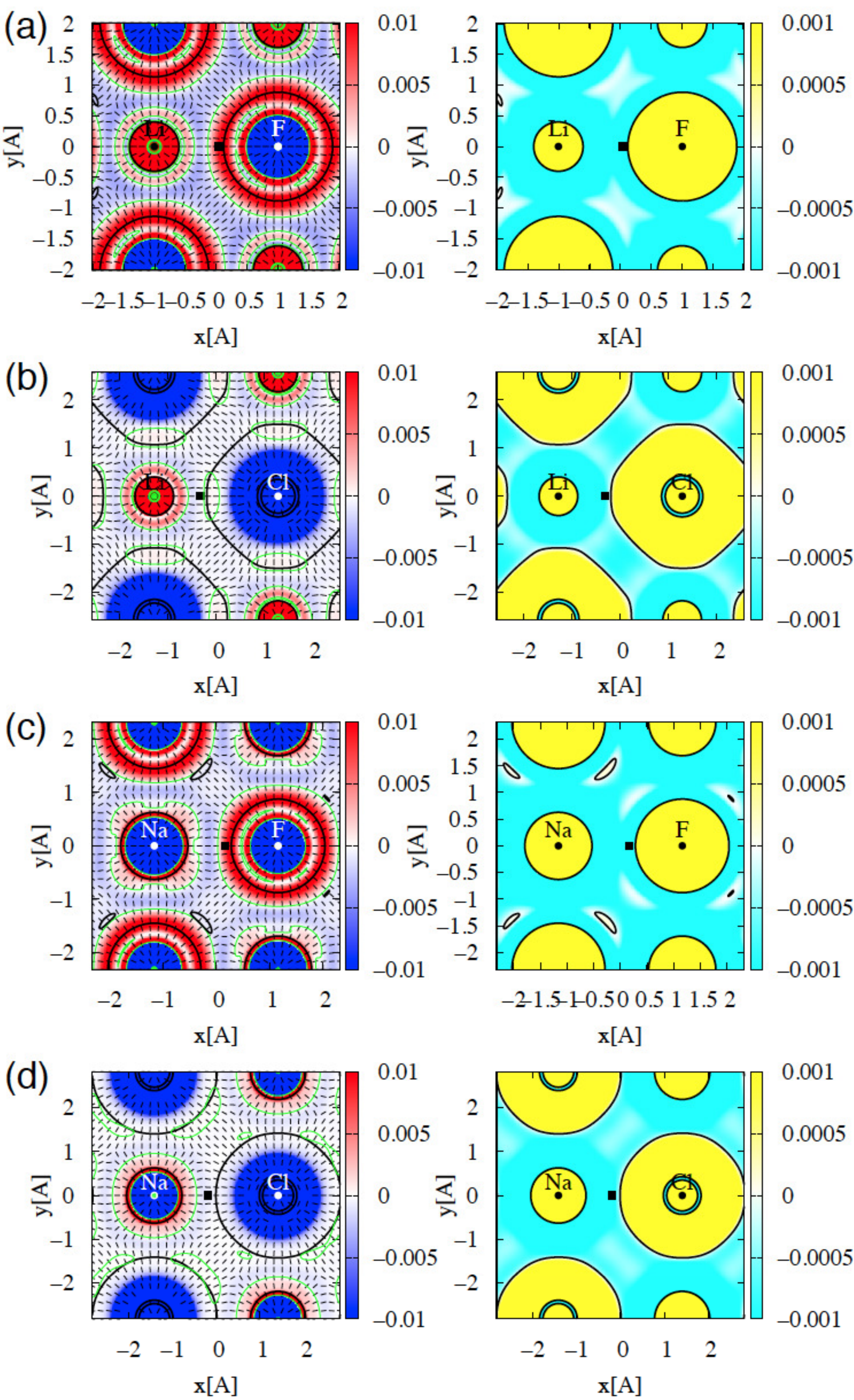}
\caption{The electronic stress tensor density (left column) and the kinetic energy density (right column) for ionic clusters (a) LiF, (b) LiCl, (c) NaF, and (d) NaCl are shown in the same manner as Fig.~\ref{fig:ionic}.
}
\label{fig:ionic}
\end{figure}

\fi


\begin{thebibliography}{99}

\bibitem{Kamaya2011}
	N.~Kamaya, K.~Homma, Y.~Yamakawa, M.~Hirayama, R.~Kanno, M.~Yonemura, T.~Kamiyama, Y.~Kato, S.~Hama, K.~Kawamoto, and A.~Mitsui, Nature Mater. {\bf 10}, 682 (2011).
	
\bibitem{Wang2015}
	Y.~Wang, W.~D.~Richards, S.~P.~Ong, L.~J.~Miara, J.~C.~Kim, Y.~Mo, and G.~Ceder, Nature Mater. {\bf 14}, 1026 (2015).

% RQED
\bibitem{Tachibana2001}
	A.~Tachibana, J. Chem. Phys. {\bf 115}, 3497 (2001).

\bibitem{Tachibana2003}
        A.~Tachibana, Field Energy Density In Chemical Reaction Systems.
        In {\it Fundamental World of Quantum Chemistry, A Tribute to the Memory of Per-Olov L\"{o}wdin}, 
        E.~J.~Br\"andas and E.~S.~Kryachko Eds., Kluwer Academic Publishers, Dordrecht (2003), Vol. II, pp 211-239. 

\bibitem{Tachibana2010}
	A.~Tachibana, J. Mol. Struct. (THEOCHEM), {\bf 943}, 138 (2010).

\bibitem{Tachibana2013}
	A.~Tachibana, Electronic Stress with Spin Vorticity. In {\it Concepts and Methods in Modern Theoretical Chemistry},
	S.~K.~Ghosh and P.~K.~Chattaraj Eds., CRC Press, Florida (2013), pp 235-251.

\bibitem{Tachibana2014a}
	A.~Tachibana, J. Comput. Chem. Jpn., {\bf 13}, 18 (2014).



%%%%%%%%%%%%%%%%
% application of RQED
\bibitem{Ikenaga2002}
	M.~Ikenaga, K.~Nakamura, A.~Tachibana and K.~Matsumoto, J. Cryst. Growth {\bf 237}, 936 (2002).

\bibitem{Hotta2002}
	S.~Hotta, K.~Doi, K.~Nakamura and A.~Tachibana, J. Chem. Phys. {\bf 117}, 142 (2002).
	
\bibitem{Hasegawa2003}
	K.~Hasegawa, K.~Doi, K.~Nakamura and A.~Tachibana, Mol. Phys. {\bf 101}, 295 (2003).

\bibitem{Yoshida2003}
	S.~Yoshida, K.~Doi, K.~Nakamura and A.~Tachibana, Applied Surface Science {\bf 216}, 141 (2003).
	
\bibitem{Makita2003}
	T.~Makita, K.~Doi, K.~Nakamura and A.~Tachibana, J. Chem. Phys. {\bf 119}, 538 (2003).

\bibitem{Tachibana2004}
	A.~Tachibana, Int. J. Quantum Chem. {\bf 100}, 981 (2004).

\bibitem{Kawakami2004}
	Y.~Kawakami, Y.~Nojima, K.~Doi, K.~Nakamura and A.~Tachibana, Electrochimica Acta {\bf 50}, 739 (2004).

\bibitem{Tachibana2005}
	A.~Tachibana, J. Mol. Model. {\bf 11}, 301 (2005).
		
\bibitem{Doi2005}
	K.~Doi, K.~Yoshida, H.~Nakano, A.~Tachibana, T.~Tanabe, Y.~Kojima and K.~Okazaki, J. Appl. Phys. {\bf 98}, 113709 (2005).

\bibitem{Nakamura2005}
	K.~Nakamura, K.~Doi, K.~Fujitani and A.~Tachibana, Phys. Rev. B {\bf 71}, 045332 (2005).

\bibitem{Nakano2006}
	H.~Nakano, H.~Ohta, A.~Yokoe, K.~Doi and A.~Tachibana, J. Power Sources, {\bf 163}, 125 (2006).

\bibitem{Szarek2007}
	P.~Szarek and A.~Tachibana, J. Mol. Model. {\bf 13}, 651 (2007).

\bibitem{Szarek2008}
	P.~Szarek, Y.~Sueda and A.~Tachibana, J. Chem. Phys. {\bf 129}, 094102 (2008).

\bibitem{Fukushima2008}
	A.~Fukushima, K.~Doi, M.~Senami and A.~Tachibana, J. Power Sources, {\bf 184}, 60 (2008).
	
\bibitem{Szarek2009}
	P.~Szarek, K.~Urakami, C.~Zhou, H.~Cheng and A.~Tachibana, J. Chem. Phys. {\bf 130}, 084111 (2009).

\bibitem{Ichikawa2009a}
%	Ichikawa, K.; Myoraku, T.; Fukushima, A.; Ishihara, Y.; Isaki, R.; Takeguchi, T.; Tachibana, A. {\it J. Mol. Struct. (THEOCHEM)} {\bf 2009}, 915, 1.
	K.~Ichikawa, T.~Myoraku, A.~Fukushima, Y.~Ishihara, R.~Isaki, T.~Takeguchi and A.~Tachibana, J. Mol. Struct. (THEOCHEM) {\bf 915}, 1 (2009).
	
\bibitem{Ichikawa2009b}
%	Ichikawa, K.; Tachibana, A. {\it Phys. Rev. A} {\bf 2009}, 80, 062507.
	K.~Ichikawa and A.~Tachibana, Phys. Rev. A {\bf 80}, 062507 (2009).

\bibitem{Ichikawa2010}
%	Ichikawa, K.; Wagatsuma, A.; Kusumoto, M.; Tachibana, A. {\it J. Mol. Struct. (THEOCHEM)} {\bf 2010}, 951, 49.
	K.~Ichikawa, A.~Wagatsuma, M.~Kusumoto and A.~Tachibana, J. Mol. Struct. (THEOCHEM), {\bf 951}, 49 (2010).

\bibitem{Fukushima2010}
	A.~Fukushima, K.~Hirai, M.~Senami and A.~Tachibana, Surface Science {\bf 604}, 1718 (2010).

\bibitem{Ichikawa2011a}
%	Ichikawa, K.; Ikeda, Y.; Wagatsuma, A.; Watanabe, K.; Szarek, P.; Tachibana, A. {\it Int. J. Quant. Chem.} {\bf 2011}, 111, 3548.
	K.~Ichikawa, Y.~Ikeda, A.~Wagatsuma, K.~Watanabe, P.~Szarek and A.~Tachibana, Int. J. Quant. Chem. {\bf 111}, 3548 (2011).

\bibitem{Ichikawa2011b}
%	Ichikawa, K.; Wagatsuma, A.; Kurokawa, Y.~I.; Sakaki, S.; Tachibana, A. {\it Theor. Chem. Acc.} {\bf 2011}, 130, 237.
	K.~Ichikawa, A.~Wagatsuma, Y.~I.~Kurokawa, S.~Sakaki and A.~Tachibana, Theor. Chem. Acc. {\bf 130}, 237 (2011).

\bibitem{Ichikawa2011c}
%	Ichikawa, K.; Wagatsuma, A.; Szarek, P.; Zhou, C.; Cheng, H.; Tachibana, A. {\it Theor. Chem. Acc.} {\bf 2011}, 130, 531.
	K.~Ichikawa, A.~Wagatsuma, P.~Szarek, C.~Zhou, H.~Cheng and A.~Tachibana, Theor. Chem. Acc. {\bf 130}, 531 (2011).

\bibitem{Fukushima2011}
	A.~Fukushima, A.~Sawairi, K.~Doi, M.~Senami, L.~Chen, H.~Cheng and A.~Tachibana, J. Phys. Soc. Jpn. {\bf 80}, 074705 (2011).
	
\bibitem{Ikeda2011}
	Y.~Ikeda, N.~Ohmori, N.~Maida, M.~Senami and A.~Tachibana, Jpn. J. Appl. Phys. {\bf 50}, 125601 (2011).

\bibitem{Senami2011}
	M.~Senami, Y.~Ikeda, A.~Fukushima and A.~Tachibana, AIP Advances {\bf 1}, 042106 (2011).
	
\bibitem{Henry2011}
	D.~J.~Henry, P.~Szarek, K.~Hirai, K.~Ichikawa, A.~Tachibana and I.~Yarovsky, J. Phys. Chem. C {\bf 115}, 1714 (2011).

\bibitem{Ichikawa2012}
	K.~Ichikawa, H.~Nozaki, N.~Komazawa and A.~Tachibana, AIP Advances {\bf 2}, 042195 (2012).

\bibitem{Nozaki2015}
	H.~Nozaki, Y.~Ikeda, K.~Ichikawa, and A.~Tachibana, J. Comput. Chem. {\bf 36}, 1240 (2015). 

\bibitem{Nozaki2015b}
	H.~Nozaki, K.~Ichikawa, and A.~Tachibana, Int. J. Quant. Chem. {\bf 116}, 504 (2016).

\bibitem{Fukuda2016}
	M.~Fukuda, K.~Soga, M.~Senami, and A.~Tachibana, Int. J. Quant. Chem. {\it in press}.

%%%%%%%%%%%%%%%%
%   Aihara
%%%%%%%%%%%%%%%%
\bibitem{Hayamizu2013}
	K.~Hayamizu and Y.~Aihara, Solid State Ionics {\bf 238}, 7 (2013).

\bibitem{Agostini2013}
	M.~Agostini, Y.~Aihara, T.~Yamada, B.~Scrosati, and J.~Hassoun, Solid State Ionics {\bf 244}, 48 (2013).

\bibitem{Hayamizu2014}
	K.~Hayamizu, Y.~Aihara, and N.~Machida, Solid State Ionics {\bf 259}, 59 (2014).

\bibitem{Ito2014}
	S.~Ito, M.~Nakakita, Y.~Aihara, T.~Uehara, and N.~Machida, J. Power Sources {\bf 271}, 342 (2014).

\bibitem{Yamada2015}
	T.~Yamada, S.~Ito, R.~Omoda, T.~Watanabe, Y.~Aihara, M.~Agostini, U.~Ulissi, J.~Hassoun, and B.~Scrosati, J. Electrochem. Soc. {\bf 162}, A646 (2015).

\bibitem{Hayamizu2015}
	K.~Hayamizu, Y.~Aihara, T.~Watanabe, T.~Yamada, S.~Ito, and N.~Machida, Solid State Ionics {\it in press}. (2015).

%%%%%%%%%%%%%%%%
%  other stress
\bibitem{Epstein1975}
%	Epstein, S.~T. {\it J. Chem. Phys.} {\bf 1975}, 63, 3573.
	S.~T.~Epstein, J. Chem. Phys. {\bf 63}, 3573 (1975).

\bibitem{Bader1980}
%	Bader, R.~F.~W. {\it J. Chem. Phys.} {\bf 1980}, 73, 2871.
	R.~F.~W.~Bader, J. Chem. Phys. {\bf 73}, 2871 (1980).

\bibitem{Bamzai1981a}
%	Bamzai, A.~S.; Deb, B.~M. {\it Rev. Mod. Phys.} {\bf 1981}, 53, 95.
	A.~S.~Bamzai and B.~M.~Deb, Rev. Mod. Phys. {\bf 53}, 95 (1981).
	
\bibitem{Nielsen1983}
%	Nielsen, O.~H.; Martin, R.~M. {\it Phys. Rev. Lett.} {\bf 1983}, 50, 697.
	O.~H.~Nielsen and R.~M.~Martin, Phys. Rev. Lett. {\bf 50}, 697 (1983).

\bibitem{Nielsen1985}
%	Nielsen, O.~H.; Martin, R.~M. {\it Phys. Rev. B} {\bf 1985}, 32, 3780.
	O.~H.~Nielsen and R.~M.~Martin, Phys. Rev. B {\bf 32}, 3780 (1985).

\bibitem{Folland1986a}
%	Folland, N.~O. {\it Phys. Rev. B} {\bf 1986}, 34, 8296.
	N.~O.~Folland, Phys. Rev. B {\bf 34}, 8296 (1986).

\bibitem{Folland1986b}
%	Folland, N.~O. {\it Phys. Rev. B} {\bf 1986}, 34, 8305.
	N.~O.~Folland, Phys. Rev. B {\bf 34}, 8305 (1986).
		
\bibitem{Godfrey1988}
%	Godfrey, M.~J. {\it Phys. Rev. B} {\bf 1988}, 37, 10176.
	M.~J.~Godfrey, Phys. Rev. B {\bf 37}, 10176 (1988).
	
\bibitem{Filippetti2000}
%	Filippetti, A.; Fiorentini, V. {\it Phys. Rev. B} {\bf 2000}, 61, 8433.
	A.~Filippetti and V.~Fiorentini, Phys. Rev. B {\bf 61}, 8433 (2000).

\bibitem{Pendas2002}
%	Pend\'{a}s, A.~M. {\it J. Chem. Phys.} {\bf 2002}, 117, 965.
	A.~M.~Pend\'{a}s, J. Chem. Phys. {\bf 117}, 965 (2002).

\bibitem{Rogers2002}
%	Rogers, C.~L.; Rappe, A.~M. {\it Phys. Rev. B} {\bf 2002} 65, 224117.
	C.~L.~Rogers and A.~M.~Rappe, Phys. Rev. B {\bf 65} 224117 (2002).

\bibitem{Morante2006}
%	Morante, S.; Rossi, G.~C.; Testa, M. {\it J. Chem. Phys.} {\bf 2006}, 125, 034101.
	S.~Morante, G.~C.~Rossi and M.~Testa, J. Chem. Phys. {\bf 125}, 034101 (2006).

\bibitem{Tao2008}
%	Tao, J.; Vignale, G.; Tokatly, I.~V. {\it Phys. Rev. Lett.} {\bf 2008}, 100, 206405.
	J.~Tao, G.~Vignale and I.~V.~Tokatly, Phys. Rev. Lett. {\bf 100}, 206405 (2008).

\bibitem{Ayers2009}
%	Ayers, P.~W.; Jenkins, S. {\it J. Chem. Phys.} {\bf 2009}, 130, 154104.
	P.~W.~Ayers and S.~Jenkins, J. Chem. Phys. {\bf 130}, 154104 (2009).

\bibitem{Jenkins2011}
%	Jenkins, S.; Kirk, S.~R.; Guevara-Garc\'{i}a, A.; Ayers, P.~W.; Echegaray, E.; Toro-Labbe, A. {\it Chem. Phys. Lett.} {\bf 2011}, 510, 18.
	S.~Jenkins, S.~R.~Kirk, A.~Guevara-Garc\'{i}a, P.~W.~Ayers, E.~Echegaray and A.~Toro-Labbe, Chem. Phys. Lett. {\bf 510}, 18 (2011).

\bibitem{GuevaraGarcia2011}
%	Guevara-Garc\'{i}a, A.; Echegaray, E.; Toro-Labbe, A.; Jenkins, S.; Kirk, S.~R.; Ayers, P.~W. {\it J. Chem. Phys.} {\bf 2011}, 134, 234106.
	A.~Guevara-Garc\'{i}a, E.~Echegaray, A.~Toro-Labbe, S.~Jenkins, S.~R.~Kirk and  P.~W.~Ayers, J. Chem. Phys. {\bf 134}, 234106 (2011).

\bibitem{Finzel2013}
	K.~Finzel and M.~Kohout, Theor. Chem. Acc. {\bf 132}, 1392 (2013).

\bibitem{Finzel2014a}
	K.~Finzel, Int. J. Quant. Chem. {\bf 114}, 568 (2014).
	
%%%%%%%%%%%
% Ehrenfest force
\bibitem{Pendas2012}
	A.~M.~Pend\'{a}s and J.~Hern\'{a}ndez-Trujillo, J. Chem. Phys. {\bf 137}, 134101 (2012).

\bibitem{Maza2013}
	J.~R.~Maza, S.~Jenkins, S.~R.~Kirk, J.~S.~M.~Anderson, and P.~W.~Ayers, Phys. Chem. Chem. Phys. {\bf 15}, 17823 (2013).
	
\bibitem{Dillen2015}
	J.~Dillen, J. Comput. Chem. {\bf 36}, 883 (2015).


%%%%%%%%%%%%%%%%%%%%%%%%%%%%%
%  WKB approximation
\bibitem{CondonShortley}
	E.~U.~Condon and G.~H.~Shortley, 
	{\it The Theory of Atomic Spectra}, Cambridge University Press, Cambridge, 1935.
	
%\bibitem{SakuraiQM}
%	J.~J.~Sakurai,
%	{\it Modern Quantum Mechanics}, Addison-Wesley Publishing Company, Inc., 


%%%%%%%%%%%%%%%%%%%%%%%%%%%%%%
%  kinetic energy density
\bibitem{Ayers2002}
	P.~W.~Ayers, R.~G.~Parr and A.~Nagy, Int. J. Quant. Chem. {\bf 90}, 309 (2002).

\bibitem{GarciaAldea2007}
	D.~Garc\'{i}a-Aldea and J.~E.~Alvarellos, J. Chem. Phys. {\bf 127}, 144109 (2007).
	
\bibitem{Anderson2010}
	J.~S.~M.~Anderson, P.~W.~Ayers and J.~I.~R.~Hernandez, J. Phys. Chem. A {\bf 114}, 8884 (2010).


%%%%%%%%%%%%
% boundary
\bibitem{Bader1987}
	R.~F.~W.~Bader, M.~T.~Carroll, J.~R.~Cheeseman, and C.~Chang, J. Am. Chem. Soc. {\bf 109}, 7968 (1987).

\bibitem{Bader}
	R.~F.~W.~Bader,
	{\it Atoms in Molecules: A Quantum Theory}, Oxford University Press, Oxford, 1990.
	
\bibitem{Popelier}
	P.~Popelier,
	{\it Atoms in Molecules; An Introduction}, Pearson Education Limited, Essex, 2000.
	







%%%%%%%%%%%%%%%%
% structures Li3PO4
%%%%%%%%%%%%%%%%
% values are taken from Du & Holzwarth 2007, which cite the paper below.

\bibitem{Keffer1967}
	C.~Keffer, A.~Mighell, F.~Mauer, H.~Swanson, and S.~Block, Inorg. Chem. {\bf 6}, 119 (1967).

\bibitem{Yakubovich1997} O.~V.~Yakubovich and V.~S.~Urusov, Crystallogr. Rep. {\bf 42}, 261 (1997).


%%%%%%%%%%%%%%%%
% structures Li3PS4
%%%%%%%%%%%%%%%%

\bibitem{Homma2011}
	K.~Homma, M.~Yonemura, T.~Kobayashi, M.~Nago, M.~Hirayama, and R.~Kanno, Solid State Ionics {\bf 182}, 53 (2011).
		
\bibitem{Mercier1982}
	R.~Mercier, J.-P.~Malugani, B.~Fahys, G.~Robert, and J.~Douglade, Acta Crystallogr. {\bf B38}, 1887 (1982).

\bibitem{Lepley2013}
	N.~D.~Lepley, N.~A.~W.~Holzwarth, and Y.~A.~Du, Phys. Rev. B {\bf 88}, 104103 (2013). 


%%%%%%%%%%%%%%%%
% structures others
%%%%%%%%%%%%%%%%

\bibitem{Kittel}
	C.~Kittel, {\it Introduction to Solid State Physics, 8th ed.}, John Wiley and Sons, Inc., New York, 2005.

\bibitem{West}
	A.~R.~West, {\it Solid State Chemistry, 2nd ed.}, John Wiley and Sons, Ltd, West Sussex, 2014.


%%%%%%%%%%%%%%%%
% computation
%%%%%%%%%%%%%%%%
%%%%%  B3LYP  %%%%%%%%%%%%
\bibitem{Lee1998}
%	Lee, C.; Yang, W.; Parr, R.~G. {\it Phys. Rev. B} {\bf 1998}, 37, 785.
	C.~Lee, W.~Yang and R.~G.~Parr, Phys. Rev. B {\bf 37}, 785 (1998).
	
\bibitem{Miehlich1989}
%	Miehlich, B.; Savin, A.; Stoll, H.; Preuss, H. {\it Chem. Phys. Lett.} {\bf 1989}, 157, 200.
	B.~Miehlich, A.~Savin, H.~Stoll and H.~Preuss,  Chem. Phys. Lett. {\bf 157}, 200 (1989).
	
\bibitem{Becke1993}
%	Becke, A.~D. {\it J. Chem. Phys.} {\bf 1993}, 98, 5648.
	A.~D.~Becke, J. Chem. Phys. {\bf 98}, 5648 (1993).	

%%%%%%%%%%%%%%%%
% Basis set
%%%%%%%%%%%%%%%%
%%%%%   6-31++G(d,p)  %%%%%%%%%%%
\bibitem{Ditchfield71}    % 6-31G for Li, O
	R.~Ditchfield, W.~J.~Hehre and J.~A.~Pople, J. Chem. Phys. {\bf 54}, 724 (1971).

\bibitem{Hehre72}  % 6-31G for O
	W.~J.~Hehre, R.~Ditchfield, and J.~A.~Pople, J. Chem. Phys. {\bf 56}, 2257 (1972).
	
\bibitem{Hariharan73} % 6-31G** 
	P.~C.~Hariharan and J.~A.~Pople, Theor. Chem. Acc. {\bf 28}, 213 (1973).

\bibitem{Francl1982}  % 6-31G** for P, S
	M.~M.~Francl, W.~J.~Pietro, W.~J~.Hehre, J.~S.~Binkley, M.~S.~Gordon, D.~J.~DeFrees, and J.~A.~Pople, J. Chem. Phys. {\bf 77}, 3654 (1982).
	
\bibitem{Frisch84}  %6-31++G, 6-311++G
	M.~J.~Frisch, J.~A.~Pople and J.~S.~Binkley, J. Chem. Phys. {\bf 80}, 3265 (1984).

%%%%%   6-311++G(d,p)  %%%%%%%%%%%
% 6-311G, 6-311G**
\bibitem{Raghavachari80b}  % 6-311G** for Li, C, F
	R.~Krishnan, J.~S.~Binkley, R.~Seeger, and J.~A.~Pople, J. Chem. Phys. {\bf 72}, 650 (1980).

\bibitem{McLean1980}  % for Na, Si, Cl
	A.~D.~McLean and G.~S.~Chandler, J. Chem. Phys. {\bf 72}, 5639 (1980).



\bibitem{Gaussian09}
Gaussian 09, Revision C.1, M. J. Frisch, G. W. Trucks, H. B. Schlegel, G. E. Scuseria, M. A. Robb, J. R. Cheeseman, G. Scalmani, V. Barone, B. Mennucci, G. A. Petersson, H. Nakatsuji, M. Caricato, X. Li, H. P. Hratchian, A. F. Izmaylov, J. Bloino, G. Zheng, J. L. Sonnenberg, M. Hada, M. Ehara, K. Toyota, R. Fukuda, J. Hasegawa, M. Ishida, T. Nakajima, Y. Honda, O. Kitao, H. Nakai, T. Vreven, J. A. Montgomery, Jr., J. E. Peralta, F. Ogliaro, M. Bearpark, J. J. Heyd, E. Brothers, K. N. Kudin, V. N. Staroverov, R. Kobayashi, J. Normand, K. Raghavachari, A. Rendell, J. C. Burant, S. S. Iyengar, J. Tomasi, M. Cossi, N. Rega, J. M. Millam, M. Klene, J. E. Knox, J. B. Cross, V. Bakken, C. Adamo, J. Jaramillo, R. Gomperts, R. E. Stratmann, O. Yazyev, A. J. Austin, R. Cammi, C. Pomelli, J. W. Ochterski, R. L. Martin, K. Morokuma, V. G. Zakrzewski, G. A. Voth, P. Salvador, J. J. Dannenberg, S. Dapprich, A. D. Daniels, \"{O}. Farkas, J. B. Foresman, J. V. Ortiz, J. Cioslowski, and D. J. Fox, Gaussian, Inc., Wallingford CT, 2009.

\bibitem{QEDynamics}
	QEDynamics, M.~Senami, K.~Ichikawa and A.~Tachibana \\
	 {\tt http://www.tachibana.kues.kyoto-u.ac.jp/qed/index.html}


\end{thebibliography}
\end{document}